\documentclass{article}
\usepackage{natbib}
\usepackage{epsfig}
\usepackage{amssymb,amsmath,tabularx}
\usepackage[T1]{fontenc}
\usepackage{color}
\usepackage{rotating}
\usepackage{array}
\usepackage{graphicx}
\usepackage{bm}
\usepackage{hhline}
\usepackage{multirow}

\newcommand{\bmath}[1]{\mbox{\boldmath $#1$}}

\begin{document}

\title{Censored Regression with Serially Correlated Errors: a
	Bayesian approach}
\author{Rodney Sousa\thanks{%
Corresponding author: rodney@ua.pt, University of Aveiro, Portugal \& CIDMA} \and %
Isabel Pereira\thanks{%
isabel.pereira@ua.pt, Departamento de Matem\'atica, Universidade de Aveiro,
Portugal \& CIDMA} \and %
Maria Eduarda Silva\thanks{%
mesilva@fep.up.pt, Faculdade de Economia, Universidade do Porto,
Portugal \& LIADD-INESC} \and Brendan McCabe\thanks{%
Brendan.Mccabe@liverpool.ac.uk, Managment School, Chatham Building Chatham
Street, University of Liverpool, L69 7ZH} }
\date{}
\maketitle

\begin{abstract}

   The problem of estimating censored linear regression models with
   autocorrelated errors arises in many environmental and social
   studies. The present work proposes a Bayesian approach to estimate
   censored regression models with AR(p) errors. The algorithm
   developed here considers the Gibbs sampler with data augmentation
   (GDA), in which, at each iteration,  both the model parameters and
   the latent variables are sampled.  The data augmentation is
   achieved from multiple sampling of the latent variables from the
   corresponding conditional distributions. A suitable variable
   transformation allows the full likelihood to be obtained.   A
   simulation study indicates that the proposed approach produces
   estimates with a high accuracy even in scenarios where the
   proportion of censored observations is large.
   
The method is further  illustrated in a real data of cloud ceiling
height, including model checking and selection for censored time
series data. 

\textbf{keywords}: \small{Censored Data, Linear Regression,
  Autocorrelation, Bayesian Analysis, Gibbs sampler, Data augmentation }
\end{abstract}

\section{Introduction}
\label{intro1}

Censored observations arise when explicit limits are placed on the
observed data and occur  in several fields including environmental
monitoring, economics, medical and social sciences. The censoring may
due to  measuring device limitations, such as  detection limits in air
pollution or mineral concentration in water, \cite{Hopke_et_al2001}). In economics, censoring occurs when constraints or regulations
are imposed, such as on observations in international trade where exports and imports are subject to trade barriers,
\cite{Zangari_Tsurumi1996}. Since the work of \cite{Buckley_James1979}
an extensive body of literature on regression analysis with censored
responses has been developed. In addition to censoring, the data often
exhibit serial correlation, leading to the adoption of dynamic
censored models.

In the time series regression context, censoring has been addressed by
several authors. The first methodological approach to estimation of
censored regressions with autocorrelated errors was proposed by
\cite{Zeger_Brookmeyer1984}, who presented the exact likelihood
function for this model. The likelihood is constructed based on blocks
of data of variable dimensions. As the block size usually increases
with the censoring rate, maximum likelihood quickly becomes
numerically intractable. Acknowledging  this issue, the authors
suggest an approximate approach based on a
pseudo-likelihood. \cite{Park_etal2007} introduced an imputation
method to estimate an ARMA model from a censored time series. The potentially censored values are imputed from random values
simulated from their conditional distribution given the observed data and the
censoring information. The resulting time series is considered complete and
may be analysed with the usual time series methods.
\cite{Mohammad2014} proposed a quasi-EM algorithm to fit ARMA
models in the presence of censoring with the particularity of treating
missing data as a special case of censoring.
\cite{Schumacher_etal2017} suggests using a Stochastic
  Approximation of the EM technique, SAEM, based on the unconditional
likelihood function of  the linear regression models with AR($p$)
errors. These authors have shown via simulations that their method yields
consistent estimates even when the proportion of censored values is
large ($\approx 40\%$). \cite{Houseman_Virji2017} proposed a Bayesian
approach to handle exposure time series data subject to left
censoring, where the autocorrelation is modelled by a spline-based
method in order to account for non-stationary autocorrelation. \cite{Wang_Chan2018} suggested a quasi-likelihood method based on a
system of equations and performed model checking based on simulated
residuals \citep{Gourieroux1987}.

The problem of estimating  regression models with autocorrelated
errors from censored observations has also been addressed in a
Bayesian framework. \cite{Zangari_Tsurumi1996} considered three
Bayesian procedures for censored regression models with AR(1)
errors. The authors derive posterior densities for the parameters of
the model  building on the work of \cite{Zeger_Brookmeyer1984}, using
Laplace approximations, a Gibbs sampler with data augmentation and a
quadrature numerical integration procedure. However, the authors found
that the Gibbs sampler using a data augmentation algorithm failed to
converge for moderate censoring percentages (10-15\%) and strongly
correlated disturbances.
Later,  
\cite{Wei_1999} considered a censored autoregression of order $p$ with
exogenous variables (censored ARX($p$)) and developed a sampling
scheme for the conditional posterior distributions of the censored
data, successfully applying the Gibbs sampler with data
augmentation. This procedure also builds on the
\cite{Zeger_Brookmeyer1984} decomposition of the likelihood.

The present work proposes a Bayesian approach to estimate censored
regression models with AR(p) errors, as it is acknowledged that the coefficients of these
models have the usual interpretation and thus are easier to explicate
in comparison with ARX models. The algorithm developed here considers
the Gibbs sampler with data augmentation (GDA), in which, at each
iteration, both the model parameters and the latent variables are
sampled. The data augmentation 
is achieved by multiple sampling of the latent variables from the corresponding
conditional distributions. The censored observations are thus replaced by
a mean of multiple samples leading to faster convergence of the algorithm and
more accurate estimates. Under data augmentation, the computation of the
likelihood function reduces to that of the likelihood of a multivariate Gaussian
sample. In time series analysis it is usual to resort to the conditional likelihood.
However, in the current situation a suitable variable transformation allows the
full likelihood to be obtained. Additionally, a procedure for model selection and
model assessment in this Bayesian framework based on data augmentation is
proposed. The relative performance of competing models can be assessed using
the Bayes factors, based on the ratio of normalising constants under each
model, referred to as evidence. A review of some commonly used
  methods of estimating the model evidence is given in
  \cite{FrielWyse2012}. The current paper further contributes to the
  literature by showing that GDA is useful for model selection
using measures of predictive performance, traditionally named
information criteria, allowing for forecast evaluation through
leave-one-out cross-validation suitable for time series
data. Empirical experiments with synthetic and real data sets indicate
that the proposed approach overcomes the 
bias introduced by the censoring even when the censoring rate is high
(40

Finally, note that attention here is restricted to left censoring in the development
of the procedure. This is, however, easily adapted and extended to the
right censoring case as shown in its application to a time series of cloud ceiling
heights, thus demonstrating the flexibility of the procedure.

The paper is organized as follows: Section \ref{sec:clr}
defines the model under study; Section \ref{sec:bayes} describes the
proposed 
Bayesian approach with data augmentation, detailing the steps all the
required steps and 
illustrates the performance of the method under three different
censorship scenarios using synthetic data sets. Section
\ref{sec:resid}
discusses model assessment when using censored data and   Section
\ref{sec:app_real_data} analyses a  time series of cloud ceiling
heights, previously analysed by \cite{Park_etal2007} and by
\cite{Schumacher_etal2017}, which was originally collected by the
National Center for Atmospheric Research (NCAR). The data consists of
716 hourly observations in San Francisco, during the month of March
1989, of which $41.7\%$ are censored. Some final remarks and possible
future extensions are given in the conclusion.

\section{Censored Linear Regression with Autocorrelated Errors}\label{sec:clr}

A latent variable $w$ is said to be left censored at $L$ if only the
values above $L$ are recorded, while the values less or equal to this
limit are reported as $L$. The observed variable $y$ is, then,
defined as 
 
	\begin{equation}\label{eq:cens_var} 
          y=\begin{cases}w, \textrm{ if } w>L\\ 
            L, \textrm{ if }  w \leq L\end{cases} 
	\end{equation} 
 
	\noindent or, equivalently, ${y=max(w,L)}$. Similarly, if $w$ is right censored, the recorded values will be ${y=min(w,L)}$. $L$ may be thought of as a detection limit.  
 
Now consider the classic linear regression model with serially
correlated errors defined as an AR($p$) process, denoted as LR-AR. The
discrete time representation of this model for the
response variable $w_t$ at time $t$ is given by 
 
\begin{align}\label{eq:lr_arp} 
		w_t &=\mathbf{x}_t\bm{\beta}+u_t \\ \nonumber 
                u_t
                    &=\rho_{1}u_{t-1}+\rho_{2}u_{t-2}+...+\rho_{p}u_{t-p}+\varepsilon_t,
   \ \  \varepsilon_t\sim N(0,\sigma_{\epsilon}^2)      \\ \nonumber
              \end{align} 
               
\noindent where $\mathbf{x}_t=(1,x_{t2},\ldots,x_{tk})$ is a $1\times
k$ vector of explanatory variables or features, $\bm{\beta}=(\beta_0,
\beta_1, \ldots, \beta_{k-1})$ is the $k \time 1$ vector of regression
coefficients, $u_t$ is a stationary AR($p$) process with Gaussian
innovations ${\varepsilon_t}$ and AR coefficients
$\bm{\rho}=(\rho_1\ldots,\rho_p),$ satisfying the usual  stationarity
conditions.  

Assume now that we observe possibly censored values
${y_t=max(w_t,L)},$ where $L$ is a known censoring limit. Then we
write the Censored Linear Regression model with AR errors, CLR-AR as  
  
\begin{align}\label{eq:clr_arp} 
                 y_t & =\max(w_t,L) \\ \nonumber
		w_t & =\mathbf{x}_t\bm{\beta}+u_t, \\ \nonumber
                u_t &
                      =\rho_{1}u_{t-1}+\rho_{2}u_{t-2}+...+\rho_{p}u_{t-p}+\varepsilon_t,
  \ \  \varepsilon_t\sim N(0,\sigma_{\varepsilon}^2)  \\ \nonumber
\end{align} 
 
Henceforward, $\mathbf{y}=(y_1,\ldots, y_T)$
represents the actual recorded values which are possibly censored
while $\mathbf{w}=(w_1,\ldots, w_T)$ denotes the corresponding latent process $w_t$, 
$\mathbf{X}$ represents the $T \times k$ matrix of the regressors   and
$\bm{\theta}=(\bm{\beta}, \bm{\rho}, \sigma_{\epsilon}^2)$  the
parameter vector.
The history of a process $\omega_t$ up to time $t$  is represented by 
$\bmath{\omega}_t=(\omega_1,\ldots, \omega_t.)$

\section{Bayesian inference with data augmentation }\label{sec:bayes}

A natural approach to inference for censored data in the Bayesian
framework is based on  the Gibbs sampler \citep{GelfandSmith1990, Casella_George1992} with data augmentation, GDA
\citep{Tanner_Wong1987, Fridley_Dixion2007, Chib1992}. This approach can be
described as a two-step procedure  in each
iteration: (i)  the (possibly) censored observations are imputed  with
values generated from a truncated conditional distribution thus
originating an augmented data set  that is considered complete; (ii)  the model parameters are generated  from their full
conditional distributions. 

First, we describe the approach to the data augmentation procedure.

\subsection{Data augmentation}

Usually the data augmentation step relies on the simulation of a 
single value for  the censored observation at each iteration, a
procedure that does not account for the variance of the truncated
distribution \citep{Hopke_et_al2001}.  In
order to overcome this problem, this work proposes a new approach to
the GDA algorithm in which the censored observation is imputed with
the mean of  several,  say $m,$ values simulated from the  truncated
distribution. Numerical studies with synthetic data, see
Section~\ref{sec:sim_study}, show that this approach, denoted by
GDA-MSM, leads to posterior distributions for the parameters with good
location and dispersion properties.

The other important issue relates to the truncated distribution from
which  we impute the (possibly) censored observations. Given a data set $\mathbf{y}=(y_1,...,y_T)$ possibly with censored
observations the augmented data set
is defined as 
defined as follows,  
 
\begin{equation}\label{eq:aug_data} 
  z_t=\begin{cases}y_t, \textrm{ if } y_t>L\\ 
    z_t\sim F(z_t\lvert\mathbf{y},\bm{\theta}, z_t\leq L), \textrm{ if }  y_t= L, 
  \end{cases} 
\end{equation} 
\noindent where $F(z_t\lvert \mathbf{y}, \bm{\theta},z_t\leq L)$ is the
truncated distribution corresponding to the censored values of the
latent variable, with support in $]-\infty,L].$ Specifically, under the Gaussian  assumption
\begin{equation}\label{eq:tr1_pdf} 
	f(z_t\lvert\mathbf{y},\bm{\theta},  z_t\leq L)={\displaystyle\frac{1}{\sigma}\times\frac{\phi\big(\frac{z_t-\mu_{t\lvert \mathbf{z}_{t-1}}}{\sigma}\big)}{\Phi\big(\frac{L-\mu_{t\lvert \mathbf{z}_{t-1}}}{\sigma}\big)}\times I_{(-\infty,L)}(z_t)}, 
\end{equation} 
with $\phi(.)$ and $\Phi(.)$ denoting, respectively, the pdf and cdf
of the standard normal distribution
and

\begin{align}
  \label{eq:mu}
 \mu_{t\lvert\, \mathbf{z}_{t-1}} = &
                                      \rho_{1}z_{t-1}+\rho_{2}z_{t-2}+...+\rho_{p}z_{t-p} \\ \nonumber
   & +(\mathbf{x}_t- \rho_{1}\mathbf{x}_{t-1}-\rho_{2}\mathbf{x}_{t-2}-...\rho_{p}\mathbf{x}_{t-p})\bm{\beta}.
  \label{eq:cond_ev1} 
  \end{align} 

The resulting vector of
augmented data $\mathbf{z}=(z_1,\ldots,z_T)$ is regarded as a
hypothetical observations of the latent variable which satisfy the
model expressed in equation (\ref{eq:lr_arp}) and is the object of ensuing
Bayesian analysis. The following sections introduce the elements
required for Bayesian analysis: likelihood and full conditional distributions.

\subsection{Complete Likelihood}

To compute the complete likelihood function  
 
\begin{equation}\label{eq:lk1} 
  L(\mathbf{z}\lvert\mathbf{y},\bm{\theta})=f(z_1, \ldots, z_T\lvert\mathbf{y}, \bm{\theta}) 
\end{equation} 
 
\noindent consider the following variable transform 
 
\begin{equation} \label{eq:transf} 
  \mathbf{z}^{\ast}=\mathbf{Q} \mathbf{z}  
\end{equation} 
 
\noindent  where the $T\times T$ matrix $\mathbf{Q}$ 
is such that $\mathbf{Q'}\mathbf{Q}$ is proportional to the inverse of the variance-covariance matrix of $\mathbf{u}.$ In fact  $\mathbf{Q}$ is a matrix of the form 
 
\begin{equation} 
  \mathbf{Q} = \left [ 
\begin{array}{ccccccccccc}  
  q_{11} & 0 & \cdots & 0 & 0 & 0& \cdots& 0&\cdots & 0& 0  \\ 
  q_{21} & q_{22} & \cdots & 0 & 0  & 0 &\cdots& 0&\cdots& 0 & 0  \\ 
  \cdots & \cdots& \cdots &\cdots & \cdots & \cdots &\cdots & \cdots & \cdots & \cdots  & \cdots\\ 
  q_{p1} & q_{p2} & \cdots & q_{pp} & 0 & 0& \cdots&0& \cdots & 0 & 0 \\ 
  -\rho_{p} & -\rho_{p-1} & \cdots & -\rho_{1} & 1 &0 &\cdots&0&\cdots & 0 & 0  \\ 
  0 & -\rho_{p} & \cdots & -\rho_{2} &  -\rho_{1}& 1&\cdots & 0&\cdots & 0 & 0  \\ 
  \cdots & \cdots& \cdots &\cdots & \cdots & \cdots & \cdots& \cdots &\cdots & \cdots  & \cdots\\ 
   0 & 0 & \cdots & 0&  0&  0& \cdots &  -\rho_{p-1}&\cdots & 1 & 0\\ 
  0 & 0 & \cdots & 0&  0& 0&\cdots &  -\rho_{p} & \cdots & -\rho_{1} & 1 \\ 
\end{array} 
\right ] 
\end{equation} 
 
\noindent with the elements $q_{ij}$ obtained under the restriction expressed in equation (\ref{eq:Q1}). 
In fact, this transform is induced by the following relationship between $\bm{\varepsilon}=(\varepsilon_1,\varepsilon_2,...,\varepsilon_T)$ and $\mathbf{u}=(u_1,u_2,...,u_T)$ in model (\ref{eq:lr_arp})  
\begin{equation} \label{eq:Q} 
  \bm{\varepsilon}=\mathbf{Q} \mathbf{u}.
\end{equation} 
 
\noindent Let $\Sigma_{\bm{\varepsilon}}=\sigma^2_{\varepsilon}
\mathbf{I}$  and $\Sigma_{\mathbf{u}}$ denote the variance-covariance
matrices of $\bm{\varepsilon}$ and $\mathbf{u},$  respectively and $\mathbf{I}$ is the  identity matrix. Since 
$\Sigma_{\bm{\varepsilon}} = \sigma^2_{\varepsilon} \mathbf{I}=\mathbf{Q} \Sigma_{\mathbf{u}} \mathbf{Q'} $ 
then $ \mathbf{Q}$ must satisfy 
\begin{equation} \label{eq:Q1} 
   \sigma^2_{\varepsilon} (\mathbf{Q'} \mathbf{Q}) ^{-1}= \Sigma_{\mathbf{u}} 
 \end{equation} 
 
 Therefore 
 
 \begin{equation}\label{eq:dify} 
	z_t^*=\begin{cases}q_{11} z_1,~~t=1\\ 
		 q_{21} z_1+q_{22} z_2,~~t=2\\ 
\vdots\\ 
       q_{p1} z_1+\ldots+q_{pp} z_p,~~t=p\\ 
		z_t-\rho_1 z_{t-1}-\ldots-\rho_p z_{t-p},~~t=p+1,...,T\end{cases} 
\end{equation} 
 
Define  $\mathbf{X}^{\ast}=\mathbf{Q} \mathbf{X}$ which results in  
 
\begin{equation}\label{eq:difx} 
	x_{tj}^*=\begin{cases}q_{11} x_{1j},~~t=1\\ 
		 q_{21} x_{1j}+q_{22} x_{2j},~~t=2\\ 
\vdots\\ 
       q_{p1} x_{1j}+\ldots+q_{pp} x_{pj},~~t=p\\ 
		x_{tj}-\rho_1 x_{t-1,j}-\ldots-\rho_p x_{t-p,j},~~t=p+1,...,T\end{cases} 
\end{equation} 
for $j=2,\ldots k$ and  $x_{t1}^*=1,~~t=1,\ldots,T.$ 
 
Then likelihood function (\ref{eq:lk1}) is  equivalent to 
 
\begin{equation}\label{eq:lk2} 
	{\displaystyle L(\mathbf{z}^*\lvert\mathbf{y},\bm{\theta})=\lvert\mathbf{Q}\lvert {\Big(\frac{1}{2\pi\sigma^2}\Big)}^{\frac{T}{2}}\exp{\Big\{-\frac{1}{2\sigma^2}\sum_{t=1}^T (z_t^*-\mathbf{x}_t^*\bm{\beta})^2\Big\}}} 
\end{equation}

\subsection{Full Conditional Distributions} 
 
Bayesian analysis involves formal consideration of prior information
and inferences about the model parameters are obtained from the
posterior distribution, $\pi(\bm{\theta}\lvert\mathbf{y})$, defined  by 
\begin{equation}\label{eq:pd_1} 
	\pi(\bm{\theta}\lvert\mathbf{y})\propto L(\mathbf{y}\lvert\bm{\theta})\times\pi(\bm{\theta}), 
\end{equation} 
where $\bm{\theta}$ is the parameters vector,
$L(\mathbf{y}\lvert\bm{\theta})$ is the likelihood function of the
observed data and $\pi(\bm{\theta})$ represents the joint prior
distribution of the parameters.
In the absence of prior information, noninformative
prior distributions are considered,  assumming that $\bm{\beta}$, $\sigma^2$ and
$\bm{\rho}$ are  independent variables with the following prior  specifications

\begin{equation}\label{eq:prior2} 
	\pi(\bm{\beta})\propto c_1, \ \ \pi(\sigma^2)\propto\frac{1}{\sigma^2}, \ \ \pi(\bm{\rho})\propto c_2\times\bm{I}_{\bm{\rho}\in S_{\bm{\rho}}}, 
\end{equation} 
where $c_1$, $c_2$ are constants, $S_{\bm{\rho}}$ is the region of stationarity of the process $u_t$ and $\bm{I}_{(\cdot)}$ denotes the indicator function. 
 
By combining (\ref{eq:lk2}) and (\ref{eq:prior2}), the posterior distribution with the augmented data is written as follows: 

\begin{equation}\label{eq:pd_2b} 
 \pi(\bm{\theta}\lvert\mathbf{y,z})\propto  \lvert\mathbf{Q}\lvert {\Big(\frac{1}{\sigma^2}\Big)}^{\frac{T}{2}+1}\exp{\Big\{-\frac{1}{2\sigma^2}\sum_{t=1}^T (z_t^*-\mathbf{x}_t^*\bm{\beta})^2\Big\}}\times\bm{I}_{\bm{\rho}\in S_{\bm{\rho}}}. 
\end{equation} 
 
From (\ref{eq:pd_2b}) it follows that the full conditional
distributions for the model parameters are given by
 
\begin{equation}\label{eq:pd_beta} 
	{\displaystyle\pi(\bm{\beta}\lvert\sigma^2,\bm{\rho},\mathbf{y,z})\propto \exp{\Big\{-\frac{1}{2}(\bm{\beta-\hat{\beta}})'\frac{1}{\sigma^2}(\mathbf{{X^*}'X^*})(\bm{\beta-\hat{\beta}})\Big\}}}, 
\end{equation} 
\begin{equation}\label{eq:pd_sigma2} 
	{\displaystyle\pi(\sigma^2\lvert\bm{\beta},\bm{\rho},\mathbf{y,z})\propto {\Big(\frac{1}{\sigma^2}\Big)}^{\frac{T}{2}+1}\exp{\Big\{-\frac{1}{2\sigma^2}\sum_{t=1}^T (z_t^*-\mathbf{x}_t^*\bm{\beta})^2\Big\}}}, 
\end{equation} 
\begin{equation}\label{eq:pd_rho} 
	{\displaystyle\pi(\bm{\rho}\lvert\bm{\beta},\sigma^2,\mathbf{y,z})\propto \lvert\mathbf{Q}\lvert \  \exp{\Big\{-\frac{1}{2\sigma^2}\sum_{t=1}^T (z_t^*-\mathbf{x}_t^*\bm{\beta})^2\Big\}}\times\bm{I}_{\bm{\rho}\in S_{\bm{\rho}}}}, 
\end{equation} 
where $\hat{\bm{\beta}}=\mathbf{({X^*}'X^*)^{-1}{X^*}'z^*}$ is the Feasible Generalized Least Squares (FGLS) estimator. The functional forms of (\ref{eq:pd_beta}) and (\ref{eq:pd_sigma2}) show that 
\begin{equation}\label{eq:pd_betab} 
	{\displaystyle\bm{\beta}\lvert\sigma^2,\bm{\rho},\mathbf{y,z}\sim N(\hat{\bm{\beta}},\sigma^2(\mathbf{{X^*}'X^*)}^{-1})},
\end{equation} 
\begin{equation}\label{eq:pd_sigma2b} 
	{\displaystyle\sigma^2\lvert\mathbf{\beta},\bm{\rho},\mathbf{y,z}\sim IG\Big(\frac{T}{2},\frac{1}{2}\mathbf{(z^*-X^*\bm{\beta})'(z^*-X^*\bm{\beta})}\Big)}. 
\end{equation} 
However, to sample values of $\bm{\rho}$ we need to use the Metropolis-Hastings algorithm within the Gibbs sampler \citep{Gilks_etal1995}.\\ 
 
\subsection{GDA-MSM algorithm}

The following algorithm describes how  to perform Bayesian inference
in the  CLR-AR($p$) using the Gibbs sampler with the described data
augmentation procedure, GDA-MSM.

Given a data set $\mathbf{y}=(y_1,...,y_T),$ possibly with censored
observations, the GDA-MSM algorithm allows the construction of a Markov Chain for the parameters of the CLR-AR($p$) model as follows: 
\begin{center}\label{tab:gda} 
	\begin{tabular}{l} 
		\hline\noalign{\smallskip} 
		Algorithm 1: Gibbs sampler with Data augmentation (GDA)\\ 
		\noalign{\smallskip}\hline\noalign{\smallskip} 
		1. Initialize with $\mathbf{y}$, $L\in\mathbf{R}$, ${N\in\mathbf{Z}}$ and $\bm{\theta}^{(0)}=(\bm{\beta}^{(0)},{\sigma^2}^{(0)},\bm{\rho}^{(0)})$\\ 
		2. Set $\mathbf{z}^{(0)}=\mathbf{y}$\\ 
		3. For $i=1,...,N$ \\ 
		4.\hspace{0.5cm} Sample $\bm{\beta}^{(i)}\sim \pi(\bm{\beta}\lvert{\sigma^2}^{(i-1)},\bm{\rho}^{(i-1)},\mathbf{y,z}^{(i-1)})$ \\ 
		5.\hspace{0.5cm} Sample ${\sigma^2}^{(i)}\sim \pi(\sigma^2\lvert\bm{\beta}^{(i)},\bm{\rho}^{(i-1)},\mathbf{y,z}^{(i-1)})$ \\ 
		6.\hspace{0.5cm} Sample $\bm{\rho}^{(i)}\sim \pi(\bm{\rho}\lvert\bm{\beta}^{(i)},{\sigma^2}^{(i)},\mathbf{y,z}^{(i-1)})$ \\ 
		7.\hspace{0.5cm} For $t=1,...,T$ \\ 
		8.\hspace{1cm} If $y_t\leq L$\\ 
          9.\hspace{1.5cm} For $j=1,...,m$ \\ 
		10.\hspace{2cm} Sample $z_{tj}\sim F(z_t\lvert\mathbf{y},\bm{\theta}^{(i)},z_t\leq L)\times I_{(z_t\leq L)}$\\ 
		11.\hspace{2cm} $z_t^{(i)}:=\displaystyle{\frac{1}{m}\sum_{j=1}^m z_{tj}}$\\ 
		12.\hspace{1cm}Else\\ 
		13.\hspace{1.5cm}$z_t^{(i)}:=y_t$\\ 
		14. Return $\mathbf{\Theta}=[\bm{\theta}^{(1)}\cdots\bm{\theta}^{(N)}]^\prime$ and $\mathbf{z}^{(N)}$.\\ 
		\hline 
	\end{tabular} 
\end{center} 
 
 The MCMC estimates of the model parameters $\bm{\theta}$ are usually
 obtained by calculating the sample mean of the GDA output
 $\mathbf{\Theta}$ , unless the marginal posterior density indicates a
 highly skewed distribution; in this case   it is more appropriate to
 use the sample median. The resulting augmented data
 $\mathbf{z}^{(N)}=(z_1^{(N)},\ldots,z_T^{(N)})$ can be regarded as
 observations on  the  latent variable for further inferences \citep{Tanner_Wong1987,Law_Jackson2017}.

\subsection{Illustration with synthetic data sets}\label{sec:sim_study}

The performance of the above procedure is illustrated  with censored
time series simulated from the CLR-AR($1$) model  with and without
explanatory varibles, several positive and negative values for the lag
 1 correlation, namely, $\rho_1=-0.8,-0.48,-0.15,0.15,0.48,0.8,$ three
 different scenarios of censorship $5\%,20\%$ and $40\%$ and three
 sample sizes $100, 500$ and $1000.$ Values for the model   parameters
 $\beta_0,\beta_1,\sigma^2$ were chosen based on the papers 
 \cite{Schumacher_etal2017} and \cite{Wang_Chan2018} and are given in Table \ref{tab:modelpar}. Note that the
 model designated as \textbf{M1} corresponds to an AR(1) with  mean
 $\beta_0/(1-\rho_1).$ The total number of models is eighteen, leading
 to $18\times 3 \times 3=162 $ simulation scenarios.  The simulation allows control of the degree of censorship and of serial correlation. 
 
\begin{table}[h!] 
	\begin{center} 
		\caption{Parameters of the CLR-AR($1$) model in the simulation study} 
		\label{tab:modelpar} 
		\begin{tabular}{cccc } 
                  Model &\multicolumn{3}{c}{Parameter}\\ \hline 
                 &  $\beta_0$ & $\beta_1$ &  $\sigma^2$ \\ \cline{2-4} 
                    \textbf{M1} &  2 & 0 & 2 \\ \hline 
                  \textbf{M2} &  2 & 1 & 2 \\ \hline 
                      \textbf{M3} & 0. 2 & 0.4 & 0.607 \\ \hline 
                                  \noalign{\smallskip}\hline 
		\end{tabular} 
	\end{center} 
\end{table}

The  procedure is implemented in R \citep{Rbase} and, in particular,
the packages $MASS$ \citep{MASS2021} and $invgamma$ 
\citep{invgamma2017} are  used to sample from the 
multivariate normal and from the inverted gamma distributions. The
algorithm is iterated $N=4\times10^4$ times, the $b=2\times10^4$
initial burn-in iterations were discarded and only every 20th value of
the last iterations is kept to reduce the autocorrelation within the
chain. The convergence of the MCMC algorithm was duly analysed with
the usual diagnostic tests available in package $Coda$
\citep{Plumer_etal2006,Robert_Casella2013}. The initial
estimates are obtained by FGLS estimates and the model parameters are
estimated by posterior means from the  remaining $M=1\times10^3$ values in
the chain. The number of simulated values for the data augmentation is
$m=5$  as used by  \cite{Hopke_et_al2001}.

\begin{figure}[h] 
	\includegraphics[scale=0.6]{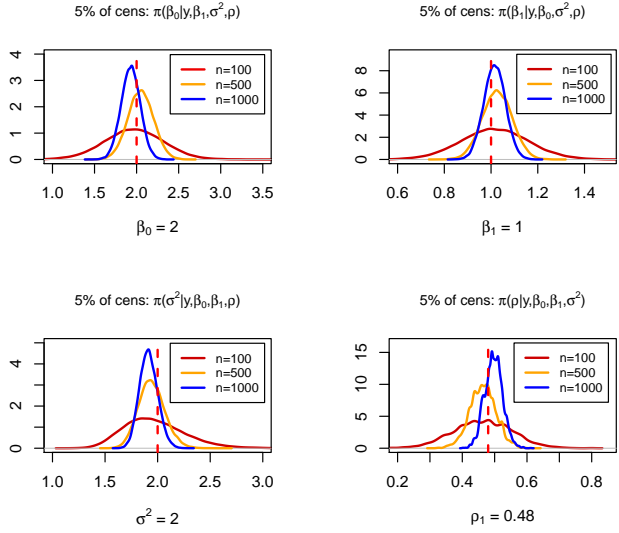} 
	\caption{Model 2 with $\rho_1=0.48$: Posterior density of the model parameters for $n=100, 500$ and $1000$ under $5\%$ of censorship.} 
	\label{fig:pd_2d_048_05} 
      \end{figure}

\begin{figure}[h] 
	\includegraphics[scale=0.6]{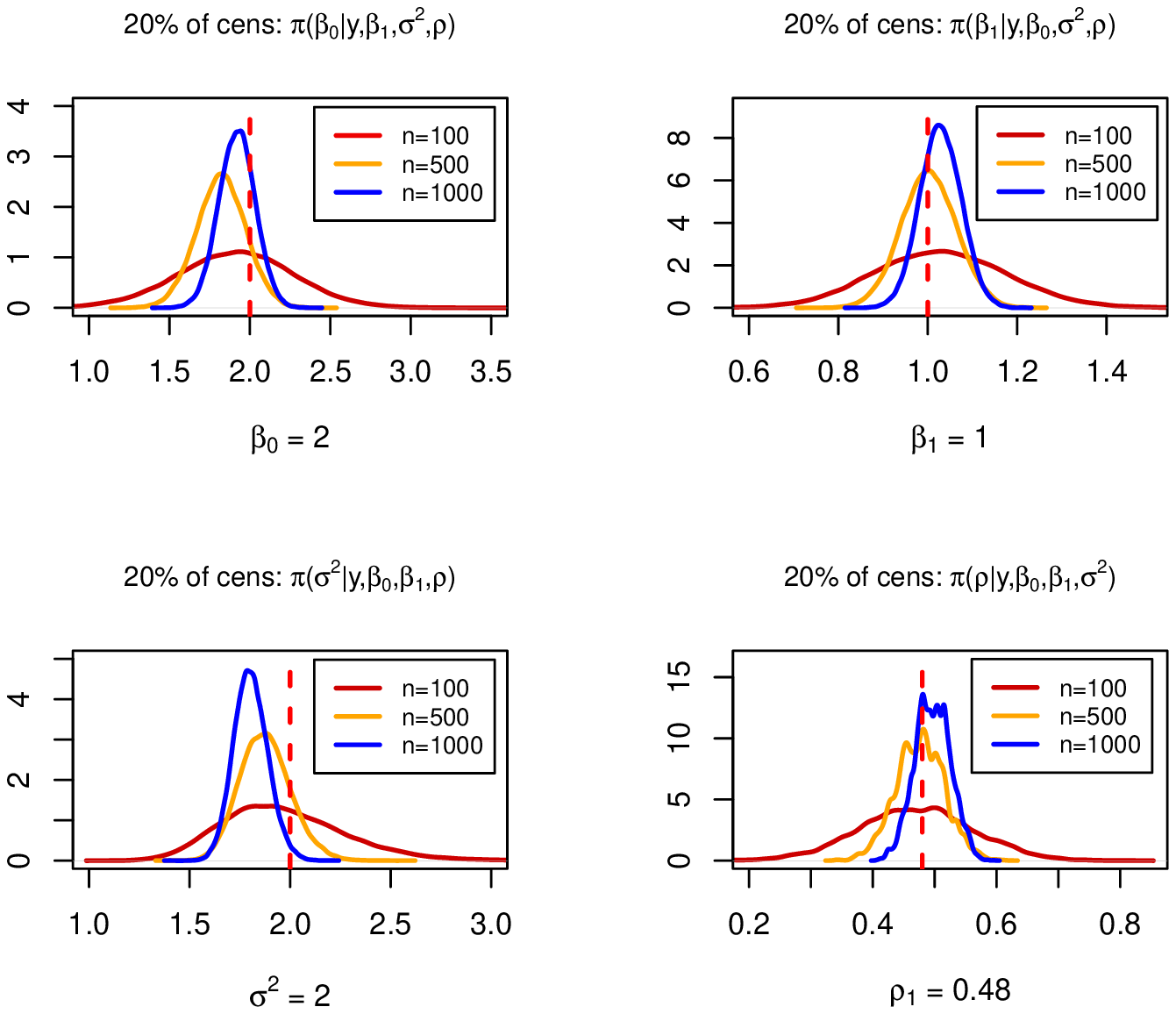} 
	\caption{Model 2 with $\rho_1=0.48$: Posterior density of the model parameters for $n=100, 500$ and $1000$ under $20\%$ of censorship.} 
	\label{fig:pd_2d_048_20} 
      \end{figure}

\begin{figure}[h] 
	\includegraphics[scale=0.6]{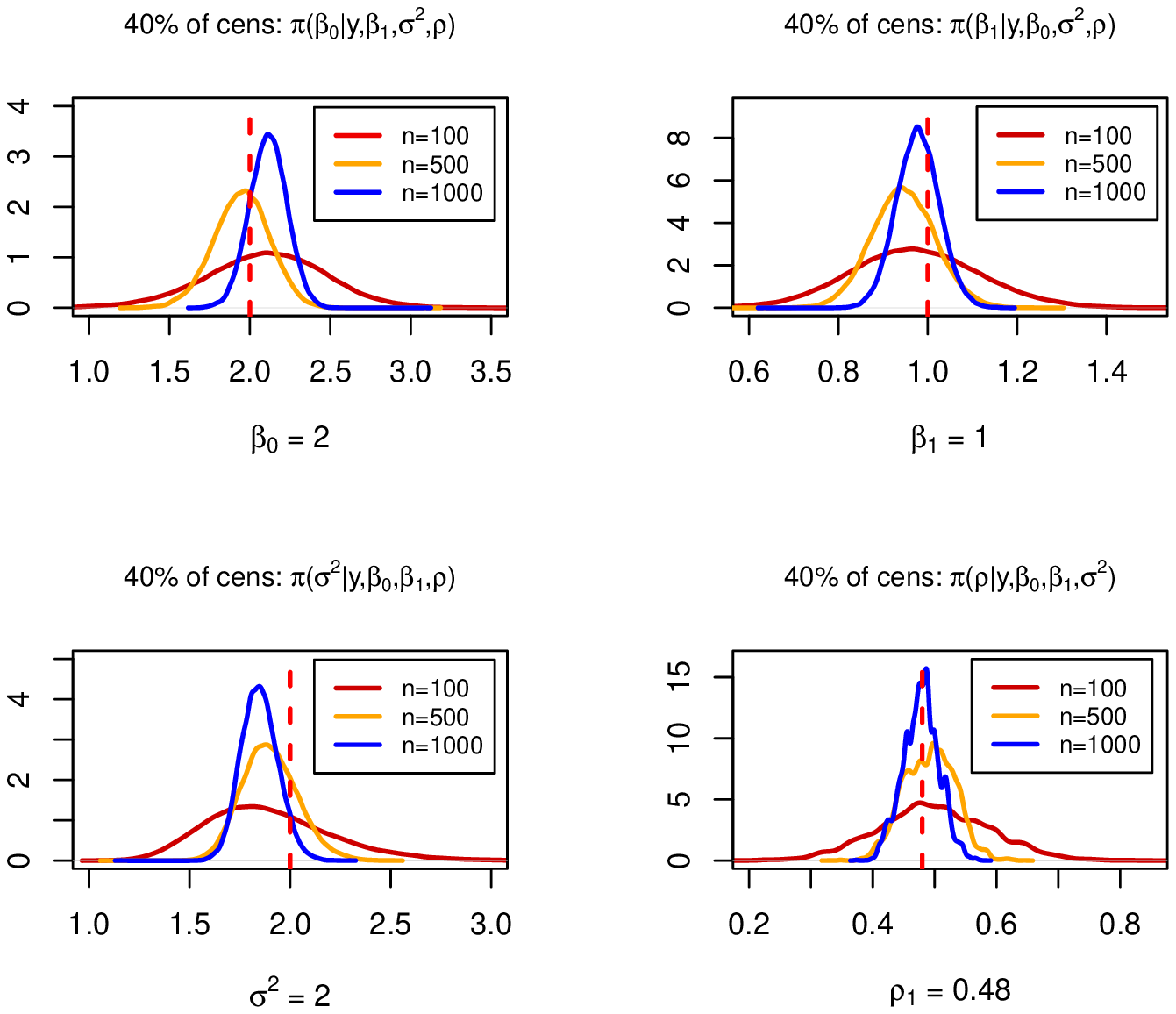} 
	\caption{Model 2 with $\rho_1=0.48$; Posterior density of the model parameters for $n=100, 500$ and $1000$ under $40\%$ of censorship.} 
	\label{fig:pd_2d_048_40} 
\end{figure}

The overall results and performance of the method are illustrated by
the posterior densities for \textbf{M2} with $\rho_1=0.48$ under the
three censorship scenarios in Figures
\ref{fig:pd_2d_048_05}--\ref{fig:pd_2d_048_40}. The plots illustrate
the efficiency and Bayes consistency of the GDA--MMS method: the
marginal posterior distributions are, in general, concentrated on sets
containing the true values of the parameters (\textit{vertical dashed
  red lines}), with the variability decreasing as the sample sizes
increase, for all the scenarios of censorship considered. Illustration
of posterior densities for other values of $\rho_1$  and the other models are presented in Appendix \ref{sec:appA}.

To further study the properties of the method, the $100$ realizations of each  the 162 scenarios is generated and the results are summarized in Tables \ref{tab:model21} to \ref{tab:model12} in Appendix \ref{sec:appB}. 
 
Thus,  the approach works well at estimating censored regression models with AR errors. 
 
\section{Model assessment in censored data}\label{sec:resid}

This section presents criteria for model assessment and model
selection in the context of Bayesian analysis of censored regressions
with autocorrelated errors.

First define  \textit{jackknife}
one-step-ahead residuals at $t+1$
\citep{Harrison_West1991,Shiffrin_etal_2008}  as 
\begin{equation}\label{eq:jres}
	{\displaystyle d_{t+1}^J=\frac{z_{t+1}-E[z_{t+1}\lvert\mathbf{z}_{t}]}{\sqrt{Var[z_{t+1}\lvert\mathbf{z}_{t}]}}}
\end{equation}
 
 \noindent  which are calculated by adopting the leave-future-out-cross-validation (LFO-CV) method \citep{Burkner_etal2020}, a modification
 of the popular leave-one-out cross validation method,  by  leaving
 out all future observations  to assess predictive
 performance in time series models. In practice, the sample
 $t=1,\ldots, T,$  is
 partitioned into a training set with $ n$ observations which grows
 continuously and a test set
 with the remaining $n-T$ observations,
 \cite{wagenmakers_etal_2006}. The value for $n$ is chosen so that
 estimation is consistent. 

 In practice, the   mean, $ E[z_{t+1}\lvert\mathbf{z}_t],$ and variance, $
 Var[z_{t+1}\lvert\mathbf{z}_t]$   in equation (\ref{eq:jres}),  are
 approximated by their sample counterparts
 using $M$ values generated from the one-step-ahead predictive distribution of $z_{t+1}$

\begin{equation}\label{eq:pred_d} 
	f(z_{t+1}\lvert\mathbf{z}_t)=\int_{\Theta} f(z_{t+1}\lvert\mathbf{z}_t,\bm{\theta})\pi(\bm{\theta}\lvert\mathbf{z}_t)d\bm{\theta}
\end{equation} 
\noindent which is not available in closed form. Therefore

\begin{equation}\label{eq:pred_e} 
  E[z_{t+1}\lvert\mathbf{z}_t]\approx \frac{1}{M}{\displaystyle\sum_{j=1}^{M} z_{t+1}^{(j)}}, 
\end{equation} 
\begin{equation}\label{eq:pred_var} 
	Var[z_{t+1}\lvert\mathbf{z}_t]\approx \frac{1}{M-1}{\displaystyle\sum_{j=1}^{M} \big(z_{t+1}^{(j)}-E[z_{t+1}\lvert\mathbf{z}_t]\big)^2}, 
\end{equation} 
where $z_{t+1}^{(j)}$ are simulated from $N(\mu,\sigma^2)$ with $\mu$
as in equation (\ref{eq:mu}), given a MCMC output $\bm{\theta}^{(j)}$
generated 
from $\pi(\bm{\theta}\lvert\mathbf{z}_t).$

 In the context of  censored data, the computation of the  residuals
 $d^J_t,$ equations (\ref{eq:jres}), (\ref{eq:pred_e}), (\ref{eq:pred_var}) and
 subsequent model assessment is achieved in a two step procedure: 

 \begin{description}
 \item[Step 1:] Given a (possibly) censored data set $y_1, \ldots,
   y_T$ fit the model via GDA-MSM algorithm and obtain an augmented
   data set, $\mathbf{z}=(z_1,\ldots,z_T)$;
   choose $n < T$
  
 \item[Step 2:] For each $t=n+1,\ldots,T$
   \begin{description}
   \item[2.1] Generate 
     $\bm{\theta}_t^{(j)}$ by applying  the GDA-MSM  algorithm to
     $(y_1, \ldots, y_t)$  and then generate $z_{t}^{(j)} \sim N(\mu,
     \sigma^2),$ $\mu$ given by equation (\ref{eq:mu}) for
     $j=1, \ldots, M$ 
    \item[2.2] Approximate the expectation and variance of the
      predictive distribution using equations (\ref{eq:pred_e}) and  (\ref{eq:pred_var})
  \item[2.3]   Regarding $\mathbf{z}=(z_1,\ldots,z_T)$ as the actually observed data, compute 
 (\ref{eq:jres}) (see
 \cite{Gourieroux1987}, \cite{Law_Jackson2017}).
  \end{description}
  \end{description}

  The standardized Bayesian residuals (\ref{eq:jres}) thus obtained
  may now be used to assess not only the quality of the fitted model
  but also to comparatively evaluate competing models in terms of
  their predictive performance via, e.g.,  the sum of squares (or of
  absolute values) in which case, models models with smaller values
  are favoured.

  Regarding model selection, the most popular   Bayesian criteria are
  the Deviance Information Criterion, DIC
  \citep{Spiegelhalter2002} and the Widely Applicable Information
  Criterion, WAIC  \citep{Watanabe2013}. Expressions for DIC and WAIC
  can be found in Appendix \ref{sec:appwaic}.  Given a MCMC output, an approximate value for $pw$  in WAIC measure (\ref{eq:waic}) is calculated by 
   
\begin{equation}\label{eq:pwb} 
	pw\approx\displaystyle{-2\sum_{t=1}^{T} \Big\{\frac{1}{M}\sum_{j=1}^{M}ln f(y_t\lvert\bm{\theta}^{(j)})-ln\Big[\frac{1}{M}\sum_{j=1}^{M}f(y_t\lvert\bm{\theta}^{(j)})\Big]\Big\}}
      \end{equation} 
       
  As in residual analysis, the augmented data set
  $\mathbf{z}=(z_1,\ldots,z_T)$ is used to evaluate the likelihood
  function $f(\mathbf{y}\lvert\theta)$, by replacing $y_t$ by
  $z_t,~~t=1,\ldots,T$,  obtained from the GDA algorithm.

\section{Analysis of cloud ceiling height time series}\label{sec:app_real_data}

  Consider the meteorological time series of cloud ceiling height,
  previously analyzed in \cite{Park_etal2007} and
  \cite{Schumacher_etal2017}. Cloud ceiling height is defined as the
  distance from the ground to the botton of a cloud and is measured in
  hundred feet. According to \cite{Park_etal2007} an accurate
  determination of the cloud ceiling height is important mainly
  because it is one of the major factors contributing to
  weather-related accidents and one of the major causes of flight
  delays. The recording device has a detection limit of $12000$ feet, so
  the observed data can be considered a right-censored time series.  
 
\begin{figure}[h!] 
		\includegraphics[width=12.7cm]{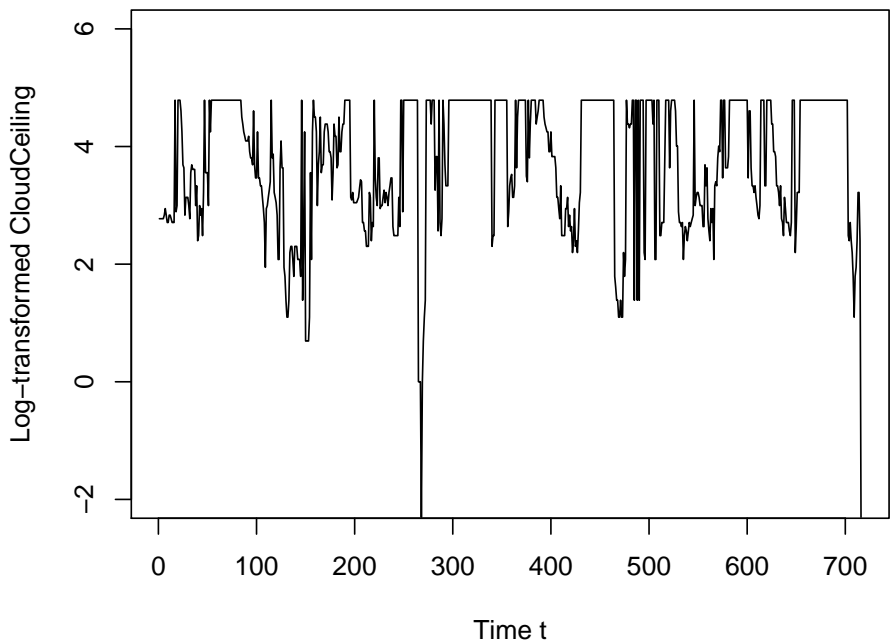} 
		\caption{Censored time series of log-transformed hourly cloud ceiling height in San Francisco during March 1989} 
		\label{fig:cloudceiling} 
\end{figure}

  The data were originally collected by the National Center for
  Atmospheric Research (NCAR) based of hourly observations in San
  Francisco, during the month of March 1989, consisting in 716
  observations, $41.7\%$ of which are censored. The log-transformed
  data is available in the package \textit{ARCensReg}
  \cite{ARCensReg2016} of the software R. A plot of the data is shown
  in the Figure \ref{fig:cloudceiling}.  
   
In the absense of explanatory variables, the  CLR-AR($p$) models
correspond to censored AR($p$) models and two values for $p=1,2$ are
considered.

\begin{figure}[h!] 
		\includegraphics[width=6cm]{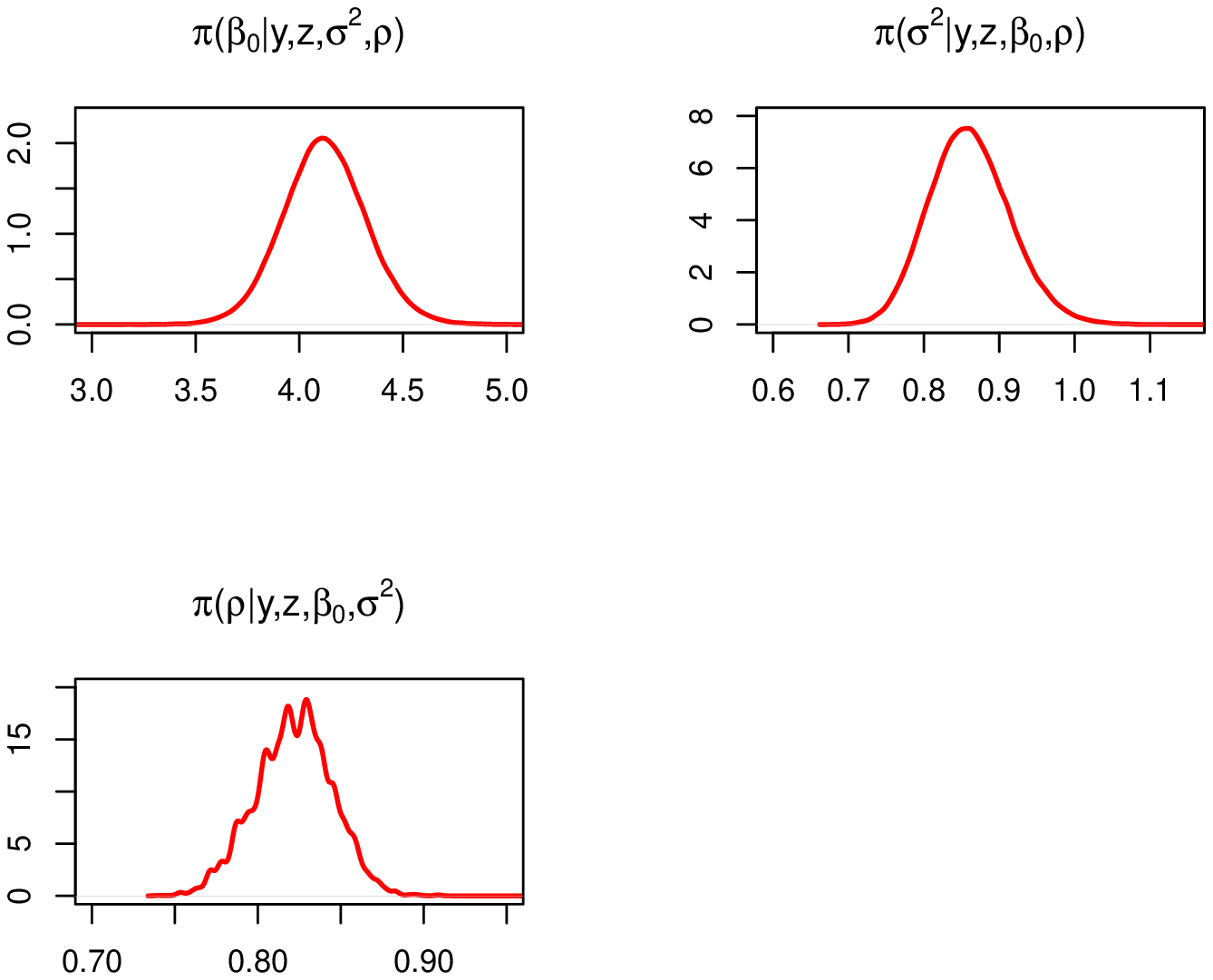} 

                \includegraphics[width=6cm]{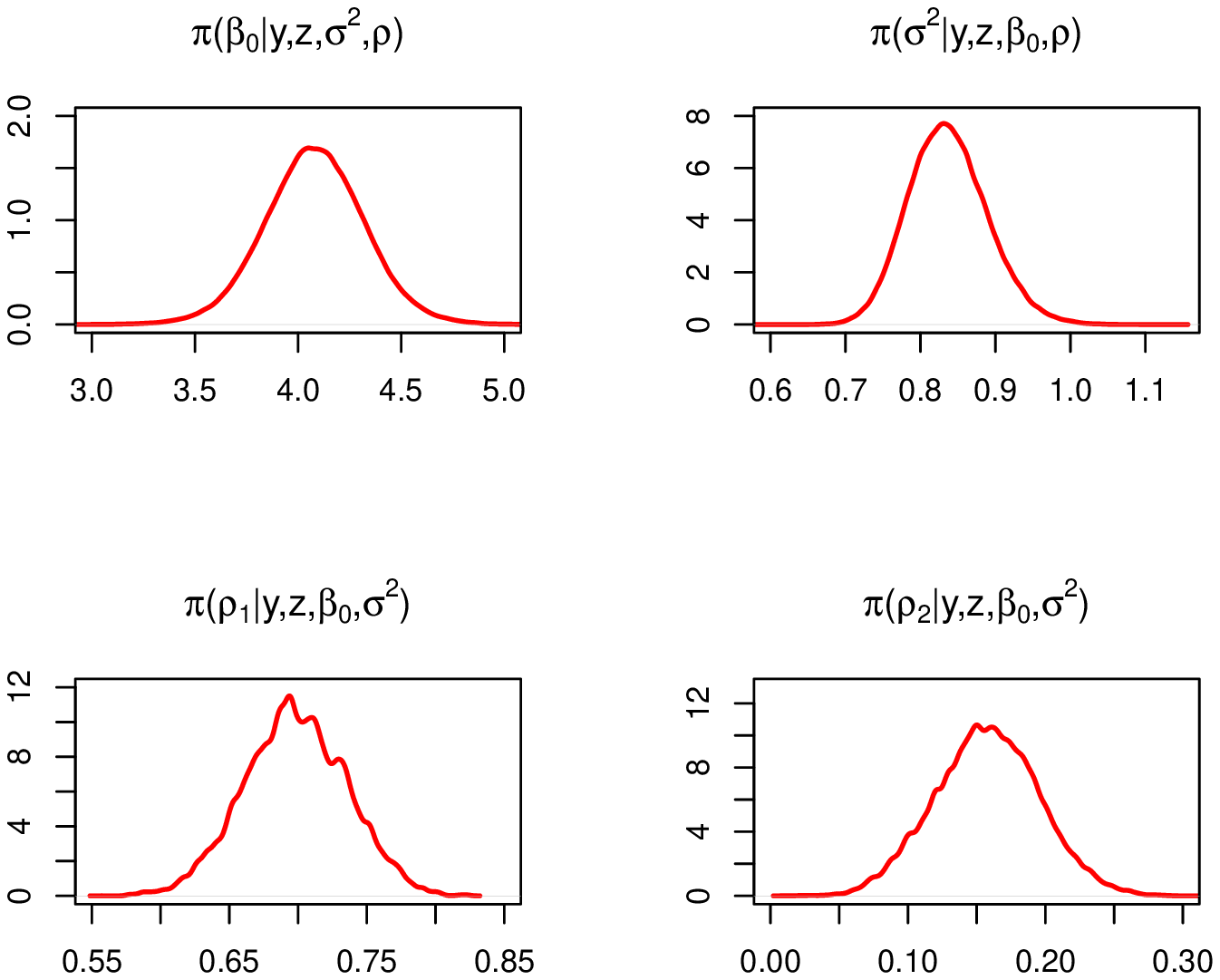} 
		\caption{Posterior densities,\textit{top}:  AR($1$) model; \textit{bottom}: AR($2$) model.} 
		\label{fig:ch5_pd_cmp} 
\end{figure}

  The burn-in is set  to $b=2\times10^4$ for
  the CLR-AR(1)  and  to $b=4\times10^4$ for the CLR-AR(2) as a result
  of monitoring the chains convergence (Appendix
  \ref{sec:convergence}).  After discarding the burn-in,  the z-scores
  of the Geweke test, \cite{Geweke1992}, are \{-0.302, 0.350, 0.624\} and \{0.272, 0.137, 0.787, -0.991\} for $p=1$ and $p=2$, respectively, suggesting the convergence of the chains (see Appendix \ref{sec:convergence}). 
  In order to reduce the autocorrelation in the MCMC outputs and get
  subsamples of length $M=1\times10^3$ to compute the estimates,
  $lag=80$  and $N=1\times10^5$ are set for $p=1$, while for $p=2$
  those values are: $lag=180$ and $N=2.2\times10^5.$ Plots of
  autocorrelation function (ACF) in Appendix \ref{sec:convergence}
  suggest no significant autocorrelation in these subsamples.  The
  resulting posterior densities for the  parameters are presented in
  Figure \ref{fig:ch5_pd_cmp}.
  
To compute the parameter estimates, the sample means of the retained
MCMC subsamples were calculated. Moreover, other summary statistics of
those subsamples are also provided in the Table \ref{tab:application},
namely, the median, the standard error (SE) and the HPD credible
interval (CI), with probability $0.95$. The CI´s were calculated using the $R$ package \textit{HDInterval} \citep{Meredith_Kruschker2018}.

\begin{figure}[h!] 
		\includegraphics[width=6cm]{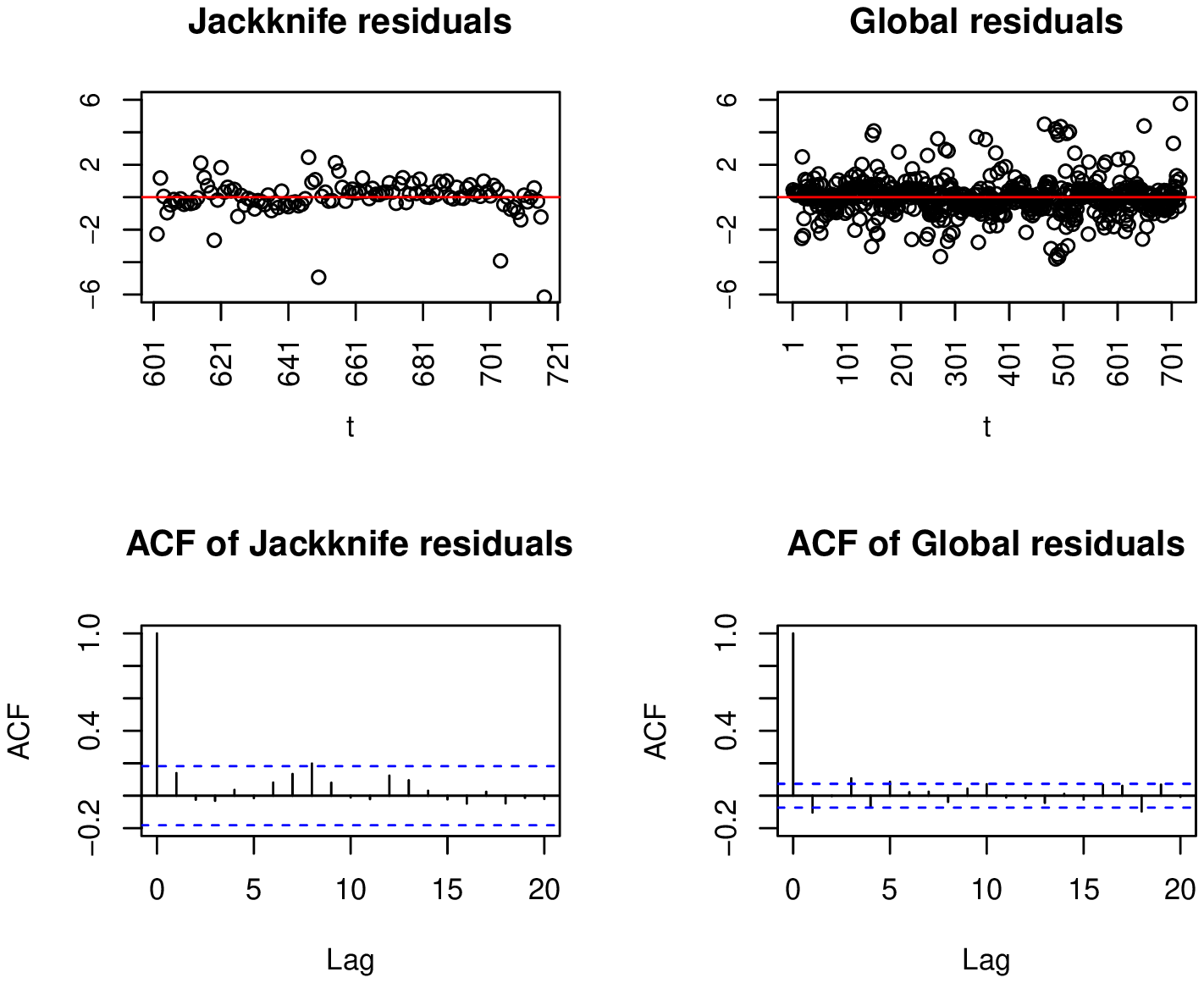} 
		\includegraphics[width=6cm]{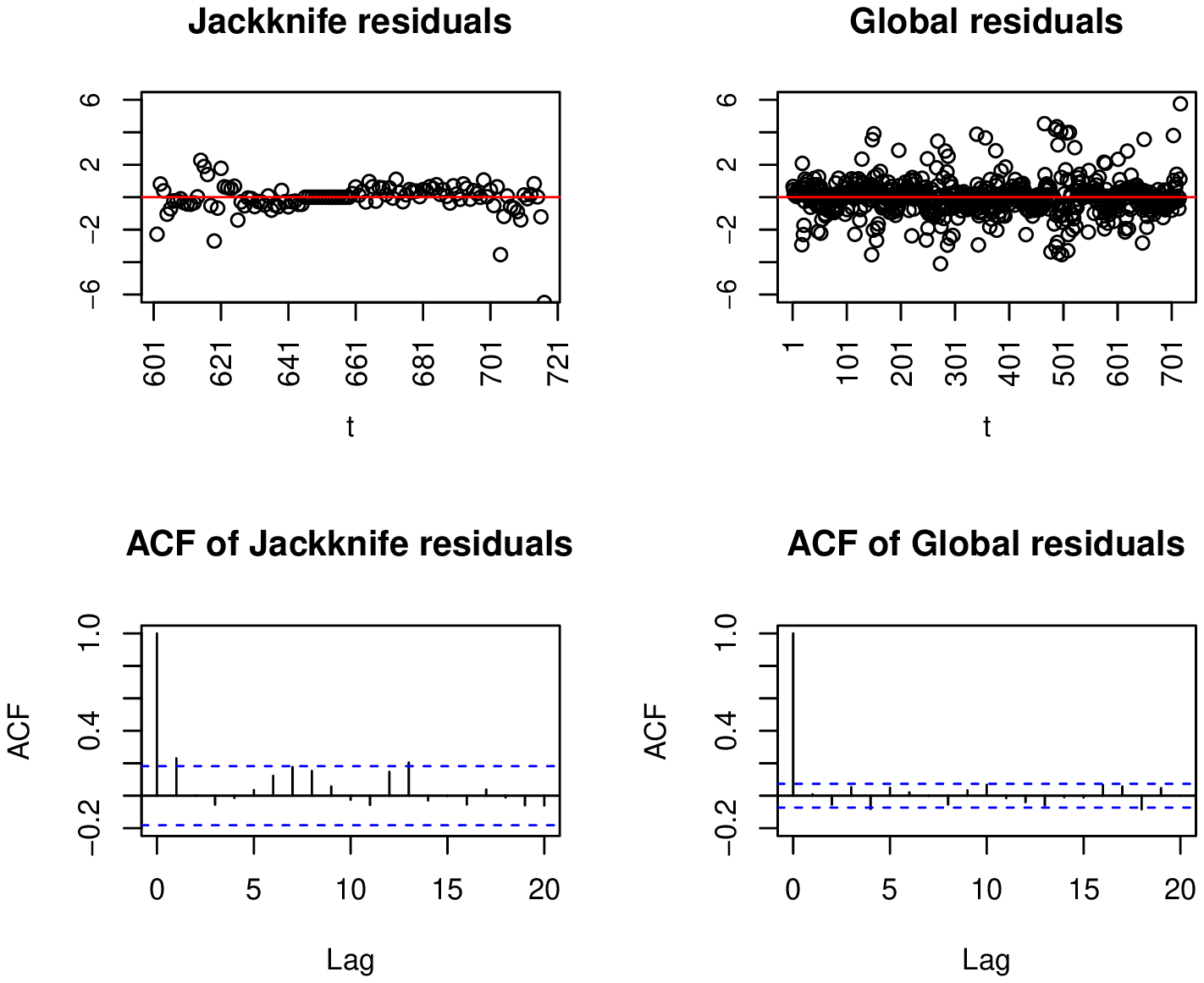} 
		\caption{(\textit{Left}) Residuals and corresponding ACF for: (\textit{Left}) $p=1$ (\textit{Right}) $p=2.$} 
		\label{fig:ch5_residual_acf} 
\end{figure}

To obtain the \textit{jackknife} forecast residuals, the size of the
initial training sample was set to $n=600$ and the corresponding mean
and variance were, respectively,  $0.007$ and $1.257$ for the AR($1$)
model, and $0.061$ and $0.95$ for the AR($2$) model. These values  are
close to $0$ and $1$, respectively, and their plots in
Figure \ref{fig:ch5_residual_acf} \textit{top} emphasize that these
residuals are distributed around zero and show no significant
correlation (Figure \ref{fig:ch5_residual_acf} \textit{bottom}). 
 
\begin{table}[h!] 
	\begin{center} 
	\caption{Parameter estimates of LR model with AR(p) error term for the log-transformed cloud ceiling height data.} 
	\label{tab:application} 
	\begin{tabular}{l l l l l l} 
	\hline\noalign{\smallskip} 
		$p$ & Stats & $\hat{\beta}_1$ & $\hat{\sigma}^2$ & $\hat{\rho}_1$ & $\hat{\rho}_2$\\ 
	\noalign{\smallskip}\hline\noalign{\smallskip} 
		\multirow{4}{*}{$1$}&Mean&4.136&0.862&0.822& -- \\ 
		&Median&4.134&0.858&0.823& -- \\ 
		&SE&0.205&0.053&0.022& -- \\ 
		&CI&[3.771, 4.543]&[0.768, 0.970]&[0.776, 0.862]& -- \\ 
	\noalign{\smallskip}\hline\noalign{\smallskip} 
		\multirow{4}{*}{\small{$\mathbf{2}$}}&Mean&\small{\textbf{4.070}}& \small{\textbf{0.839}}& \small{\textbf{0.697}}&\small{\textbf{0.160}}\\ 
		&Median&4.068&0.836&0.695&0.160\\ 
		&SE&0.248&0.053&0.037&0.037\\ 
		&CI&[3.620, 4.576]&[0.738, 0.942]&[0.626, 0.766]&[0.087, 0.231]\\ 
	\noalign{\smallskip}\hline 
	\end{tabular} 
	\end{center} 
\end{table} 
 
The values of DIC, WAIC and the sum of squared standardized jackknife residuals (SSJR) are given in Table \ref{tab:dic}. Since the model with AR($2$) error presents lowest values of DIC, WAIC and SSJR, the AR($2$) model is the chosen one. This conclusion and the values of model parameters are identical to that obtained by \cite{Schumacher_etal2017} when analysing this dataset. 
 
\begin{table}[h!] 
	\begin{center} 
		\caption{Information criteria for model assessment} 
		\label{tab:dic} 
		\begin{tabular}{cccc } 
             &\multicolumn{3}{r}{}\\ \hline 
                  Model & DIC & WAIC &  SSJR \\ \hline 
                 CLR-AR(1) &  $684.1$ & $416497.4$ & $144.6$ \\
            CLR-AR(2) &  $590.6$ &  $377259$& $109.8$ \\
                                  \noalign{\smallskip}\hline 
		\end{tabular} 
	\end{center} 
\end{table}

\begin{figure}[h!] 
		\includegraphics[width=11.7cm]{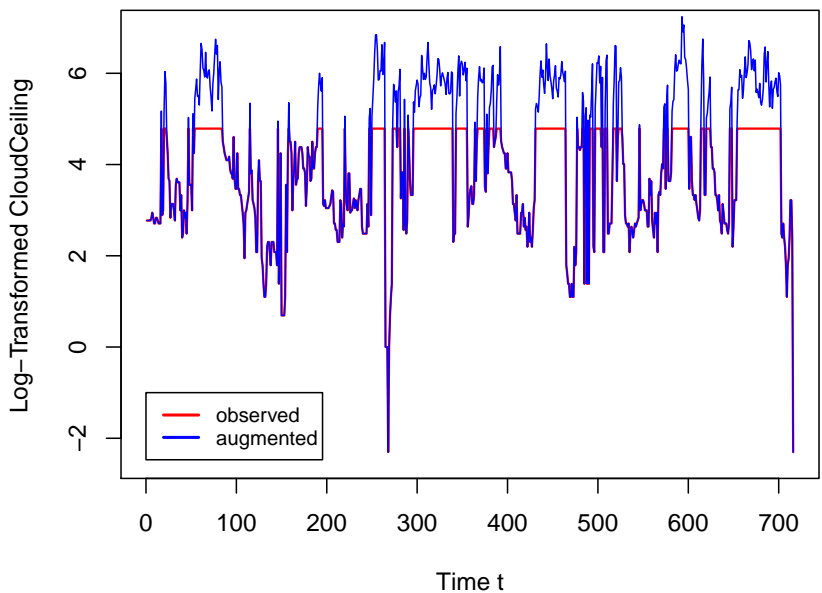} 
		\caption{Observed vs augmented data of censored time series of log-transformed hourly cloud ceiling height in San Francisco during March 1989.} 
		\label{fig:cloudCeiling_augmented} 
\end{figure}

  The augmented data, corresponding to the estimated model with AR($2$) errors is represented in Figure \ref{fig:cloudCeiling_augmented} (\textit{blue line}) against the observed data (\textit{red line}). 
 
\section{Conclusions}\label{sec:conclusion}

This work proposes a Bayesian approach to perform inference in a
linear regression  model with AR($p$) errors for censored data
(CLR-AR($p$) model). Ignoring the censorship pattern in the data and
applying  usual estimation methods results in biased estimates. The algorithm proposed implements a  
Gibbs sampler with data augmentation. The novelty stems from
the data augmentation with the mean of multiple simulations (GDA-MMS),
which improves the accuracy of the algorithm.  In fact, the GDA-MMS algorithm works well even when the proportion of censored values is large (40\%). 
 
Note that in the simulation and empirical example the Jeffrey priors
were used. However, if information about the data is available other
priors with appropriate hyperparameters may be used, in particular  for $(\bm{\beta}, \sigma^2)$ a Multivariate Normal - Inverted Gamma distribution may be considered. 
 
Here the censoring threshold was considered known. An open issue to be considere in future work is to model the data under unknown censoring level.

\section*{Acknowledgements}

{\small { This work is supported by Funda\c{c}\~{a}o  Calouste Gulbenkian and
  the Center for Research and Development  in Mathematics and
  Applications (CIDMA) through the Portuguese  Foundation for Science
  and Technology  (FCT - Funda\c{c}\~{a}o para a Ci\^{e}ncia e a
    Tecnologia),  reference UIDB/04106/2020.}
}


\begin{thebibliography}{33}
\expandafter\ifx\csname natexlab\endcsname\relax\def\natexlab#1{#1}\fi
\expandafter\ifx\csname url\endcsname\relax
  \def\url#1{\texttt{#1}}\fi
\expandafter\ifx\csname urlprefix\endcsname\relax\def\urlprefix{URL }\fi
\providecommand{\eprint}[2][]{\url{#2}}
\providecommand{\bibinfo}[2]{#2}
\ifx\xfnm\relax \def\xfnm[#1]{\unskip,\space#1}\fi
\bibitem[{Beach and MacKinnon(1978)}]{Beach_MacKinnon1978b}
\bibinfo{author}{Beach, C.}, \bibinfo{author}{MacKinnon, J.},
  \bibinfo{year}{1978}.
\newblock \bibinfo{title}{Full maximum likelihood estimation of second order
  autoregressive errors models}.
\newblock \bibinfo{journal}{Journal of Econometrics} \bibinfo{volume}{7},
  \bibinfo{pages}{187--198}.
\bibitem[{Buckley and James(1979)}]{Buckley_James1979}
\bibinfo{author}{Buckley, J.}, \bibinfo{author}{James, I.},
  \bibinfo{year}{1979}.
\newblock \bibinfo{title}{Linear regression with censored data}.
\newblock \bibinfo{journal}{Biometrika} \bibinfo{volume}{66, No. 3},
\bibinfo{pages}{429--436}.
 \bibitem[{Burkner et~al(2020)}]{Burkner_etal2020}
\bibinfo{author}{Burkner,P-C.}, \bibinfo{author}{Gabry, J.},
\bibinfo{author}{Vehtari, A}. \bibinfo{year}{2020}
\newblock \bibinfo{title}{ Approximate leave-future-out cross-validation for Bayesian time series models}.
\newblock \bibinfo{journal}{ Journal of Statistical Computation and Simulation} \bibinfo{volume}{90, No. 14}, \bibinfo{pages}{2499--2523}.
\bibitem[{Casella and George(1992)}]{Casella_George1992}
\bibinfo{author}{Casella, G.}, \bibinfo{author}{George, E.},
  \bibinfo{year}{1992}.
\newblock \bibinfo{title}{Explaining the gibbs sampler}.
\newblock \bibinfo{journal}{The American Statistician} \bibinfo{volume}{46, No.
  3}, \bibinfo{pages}{167--174}.
\bibitem[{Chib(1992)}]{Chib1992}
\bibinfo{author}{Chib, S.}, \bibinfo{year}{1992}.
\newblock \bibinfo{title}{Bayes inference in the tobit censored regression
  model}.
\newblock \bibinfo{journal}{Journal of Econometrics} \bibinfo{volume}{51},
  \bibinfo{pages}{79--99}.
\bibitem[{Fridley and Dixon(2007)}]{Fridley_Dixion2007}
\bibinfo{author}{Fridley, B.}, \bibinfo{author}{Dixon, P.},
  \bibinfo{year}{2007}.
\newblock \bibinfo{title}{Data augmentation for a bayesian spatial model
  involving censored observations}.
\newblock \bibinfo{journal}{Environmetrics} \bibinfo{volume}{18},
\bibinfo{pages}{107-–123}.
\bibitem[{Friel and Wyse(2012)}]{FrielWyse2012}
\bibinfo{author}{Friel, N.}, \bibinfo{author}{Wyse, J.}, \bibinfo{year}{2012}. \newblock \bibinfo{title}{Estimating the statistical evidence -- a review.}.
\newblock \bibinfo{journal}{Statistica Neerlandica} \bibinfo{volume}{66},
\bibinfo{pages}{288--308}. 
\bibitem[{Gelfand and Smith(1990)}]{GelfandSmith1990}
\bibinfo{author}{Gelfand, A.E.}, \bibinfo{author}{Smith, A.F.M.},
  \bibinfo{year}{1990}.
\newblock \bibinfo{title}{Sampling-based approaches to calculating marginal
  densities}.
\newblock \bibinfo{journal}{Journal of the American Statistical Association}
  \bibinfo{volume}{85, No. 410}, \bibinfo{pages}{398--409}.
\bibitem[{Geweke(1992)}]{Geweke1992}
\bibinfo{author}{Geweke, J.}, \bibinfo{year}{1992}.
\newblock \bibinfo{title}{Evaluating the accuracy of sampling-based approaches
  to the calculation of posterior moments (with discussion)}, in:
  \bibinfo{editor}{Bernardo, J.}, \bibinfo{editor}{Berger, J.},
  \bibinfo{editor}{Dawid, A.}, , \bibinfo{editor}{Smith, A.} (Eds.),
  \bibinfo{booktitle}{Bayesian Statistics 4}. \bibinfo{publisher}{Oxford
  University Press}, \bibinfo{address}{Oxford}, pp. \bibinfo{pages}{169--193}.
\bibitem[{Gilks et~al.(1995)Gilks, Best and Tan}]{Gilks_etal1995}
\bibinfo{author}{Gilks, W.R.}, \bibinfo{author}{Best, N.G.},
  \bibinfo{author}{Tan, K.K.C.}, \bibinfo{year}{1995}.
\newblock \bibinfo{title}{Adaptative rejection metroplis sampling within gibbs
  sampling}.
\newblock \bibinfo{journal}{Journal of the Royal Statistical Society}
  \bibinfo{volume}{44, 4}, \bibinfo{pages}{455--472}.
\bibitem[{Gourieroux et~al.(1987)Gourieroux, Monfort, Renault and
  Trognon}]{Gourieroux1987}
\bibinfo{author}{Gourieroux, C.}, \bibinfo{author}{Monfort, A.},
  \bibinfo{author}{Renault, E.}, \bibinfo{author}{Trognon, A.},
  \bibinfo{year}{1987}.
\newblock \bibinfo{title}{Simulated residuals}.
\newblock \bibinfo{journal}{Journal of Econometrics} \bibinfo{volume}{34},
  \bibinfo{pages}{201--252}.
\bibitem[{Harrison and West(1991)}]{Harrison_West1991}
\bibinfo{author}{Harrison, J.}, \bibinfo{author}{West, M.},
  \bibinfo{year}{1991}.
\newblock \bibinfo{title}{Dynamic linear model diagnostics}.
\newblock \bibinfo{journal}{Biometrika} \bibinfo{volume}{78, 4},
  \bibinfo{pages}{797--808}.
\bibitem[{Hopke et~al.(2001)Hopke, Liu and Rubin}]{Hopke_et_al2001}
\bibinfo{author}{Hopke, P.}, \bibinfo{author}{Liu, C.}, \bibinfo{author}{Rubin,
  D.}, \bibinfo{year}{2001}.
\newblock \bibinfo{title}{Multiple imputation for multivariate data with
  missing and below-threshold measurements: Time-series concentrations of
  pollutants in the arctic}.
\newblock \bibinfo{journal}{Biometrics} \bibinfo{volume}{57},
  \bibinfo{pages}{22–33}.
\bibitem[{Houseman and Virji(2017)}]{Houseman_Virji2017}
\bibinfo{author}{Houseman, E.A.}, \bibinfo{author}{Virji, M.A.},
  \bibinfo{year}{2017}.
\newblock \bibinfo{title}{A bayesian approach for summarizing and modeling
  time-series exposure data with left censoring}.
\newblock \bibinfo{journal}{Annals of Work Exposures and Health}
  \bibinfo{volume}{61, No. 7}, \bibinfo{pages}{773--–783}.
\bibitem[{Kahle and Stamey(2017)}]{invgamma2017}
\bibinfo{author}{Kahle, D.}, \bibinfo{author}{Stamey, J.},
  \bibinfo{year}{2017}.
\newblock \bibinfo{title}{R package 'invgamma': The inverse gamma
  distribution}.
\newblock \bibinfo{journal}{CRAN Repository} .
\bibitem[{Law and Jackson(2017)}]{Law_Jackson2017}
\bibinfo{author}{Law, M.}, \bibinfo{author}{Jackson, D.}, \bibinfo{year}{2017}.
\newblock \bibinfo{title}{Residual plots for linear regression models with
  censored outcome data: A refined method for visualizing residual
  uncertainty}.
\newblock \bibinfo{journal}{Communications in Statistics - Simulation and
  Computation} \bibinfo{volume}{46:4}, \bibinfo{pages}{3159--3171}.
\bibitem[{Meredith and Kruschke(2018)}]{Meredith_Kruschker2018}
\bibinfo{author}{Meredith, M.}, \bibinfo{author}{Kruschke, J.},
  \bibinfo{year}{2018}.
\newblock \bibinfo{title}{Package hdinterval}.
\newblock \bibinfo{journal}{CRAN Repository} , \bibinfo{pages}{1--7}.
\bibitem[{Mohammad(2014)}]{Mohammad2014}
\bibinfo{author}{Mohammad, N.M.}, \bibinfo{year}{2014}.
\newblock \bibinfo{title}{Censored Time Series Analysis}.
\newblock \bibinfo{publisher}{Phd Thesis. The University of Western Ontario},
  \bibinfo{address}{Ontario}.
\bibitem[{Park et~al.(2007)Park, Genton and Ghosh}]{Park_etal2007}
\bibinfo{author}{Park, J.}, \bibinfo{author}{Genton, M.},
  \bibinfo{author}{Ghosh, S.}, \bibinfo{year}{2007}.
\newblock \bibinfo{title}{Censored time series analysis with autoregressive
  moving average models}.
\newblock \bibinfo{journal}{The Canadian Journal of Statistics}
  \bibinfo{volume}{35, 1}, \bibinfo{pages}{151--168}.
\bibitem[{Plummer et~al.(2006)Plummer, Best, Cowles and
  Vines}]{Plumer_etal2006}
\bibinfo{author}{Plummer, M.}, \bibinfo{author}{Best, N.},
  \bibinfo{author}{Cowles, K.}, \bibinfo{author}{Vines, K.},
  \bibinfo{year}{2006}.
\newblock \bibinfo{title}{Coda: Convergence diagnosis and output analysis for
  mcmc}.
\newblock \bibinfo{journal}{R News} \bibinfo{volume}{6},
  \bibinfo{pages}{7--11}.
\bibitem[{Prais and Winsten(1954)}]{Prais_Winsten1954}
\bibinfo{author}{Prais, S.}, \bibinfo{author}{Winsten, C.},
  \bibinfo{year}{1954}.
\newblock \bibinfo{title}{Trend estimators and serial correlation}.
\newblock \bibinfo{journal}{Cowles Comission Discussion Paper: Statistics, No.
  383} .
\bibitem[{{R Core Team}(2020)}]{Rbase}
\bibinfo{author}{{R Core Team}}, \bibinfo{year}{2020}.
\newblock \bibinfo{title}{R: A Language and Environment for Statistical
  Computing}.
\newblock \bibinfo{organization}{R Foundation for Statistical Computing}.
  \bibinfo{address}{Vienna, Austria}.
\bibitem[{Ripley et~al.(2021)Ripley, Venables, Hornik, Gebhardt and
  Firth}]{MASS2021}
\bibinfo{author}{Ripley, B.}, \bibinfo{author}{Venables, B.},
  \bibinfo{author}{Hornik, K.}, \bibinfo{author}{Gebhardt, A.},
  \bibinfo{author}{Firth, D.}, \bibinfo{year}{2021}.
\newblock \bibinfo{title}{R package 'mass': Support functions and datasets for
  venables and ripley's mass}.
\newblock \bibinfo{journal}{CRAN Repository} .
\bibitem[{Robert and Casella(2010)}]{Robert_Casella2013}
\bibinfo{author}{Robert, C.}, \bibinfo{author}{Casella, G.},
  \bibinfo{year}{2010}.
\newblock \bibinfo{title}{Introducing Monte Carlo Methods with R}.
\newblock \bibinfo{publisher}{Springer}, \bibinfo{address}{New York}.
\bibitem[{Schumacher et~al.(2016)Schumacher, Lachos and
  Galarza}]{ARCensReg2016}
\bibinfo{author}{Schumacher, F.}, \bibinfo{author}{Lachos, V.},
  \bibinfo{author}{Galarza, C.}, \bibinfo{year}{2016}.
\newblock \bibinfo{title}{R package 'arcensreg': Fitting univariate censored
  linear regression model with autoregressive errors}.
\newblock \bibinfo{journal}{CRAN Repository} .
\bibitem[{Schumacher et~al.(2017)Schumacher, Lachos and
  Dey}]{Schumacher_etal2017}
\bibinfo{author}{Schumacher, F.L.}, \bibinfo{author}{Lachos, V.},
  \bibinfo{author}{Dey, D.}, \bibinfo{year}{2017}.
\newblock \bibinfo{title}{Censored models with autoregressive errors: A
  likelihood-based perspective}.
\newblock \bibinfo{journal}{The Canadian Journal of Statistics}
  \bibinfo{volume}{45, 68}, \bibinfo{pages}{375--392}.
\bibitem[{Shiffrin et~al.(2008)Shiffrin, Lee, Kim and
  Wagenmakers}]{Shiffrin_etal_2008}
\bibinfo{author}{Shiffrin, R.M.}, \bibinfo{author}{Lee, M.D.},
  \bibinfo{author}{Kim, W.}, \bibinfo{author}{Wagenmakers, E.J.},
  \bibinfo{year}{2008}.
\newblock \bibinfo{title}{A survey of model evaluation approaches with a
  tutorial on hierarchical bayesian methods}.
\newblock \bibinfo{journal}{Cognitive Science} \bibinfo{volume}{32},
  \bibinfo{pages}{1248--1284}.
\bibitem[{Spiegelhalter et~al.(2002)Spiegelhalter, Best, Carlin and van~der
  Linde}]{Spiegelhalter2002}
\bibinfo{author}{Spiegelhalter, D.J.}, \bibinfo{author}{Best, N.G.},
  \bibinfo{author}{Carlin, B.P.}, \bibinfo{author}{van~der Linde, A.},
  \bibinfo{year}{2002}.
\newblock \bibinfo{title}{Bayesian measures of model complexity and fit}.
\newblock \bibinfo{journal}{Royal Statistical Society} \bibinfo{volume}{14},
  \bibinfo{pages}{867--897}.
\bibitem[{Tanner and Wong(1978)}]{Tanner_Wong1987}
\bibinfo{author}{Tanner, M.}, \bibinfo{author}{Wong, W.}, \bibinfo{year}{1978}.
\newblock \bibinfo{title}{The calculation of posterior distributions by data
  augmentation}.
\newblock \bibinfo{journal}{Journal of American Statistical Association}
  \bibinfo{volume}{82, No. 398}, \bibinfo{pages}{528--540}.
\bibitem[{Wagenmakers et~al.(2006)Wagenmakers, Gruwald and
  Steyvers}]{wagenmakers_etal_2006}
\bibinfo{author}{Wagenmakers, E.J.}, \bibinfo{author}{Gruwald, P.},
  \bibinfo{author}{Steyvers, M.}, \bibinfo{year}{2006}.
\newblock \bibinfo{title}{Accumulative prediction error and selection of time
  series models}.
\newblock \bibinfo{journal}{Journal of Mathematical Psycology}
  \bibinfo{volume}{50}, \bibinfo{pages}{149--166}.
\bibitem[{Wang and Chan(2018)}]{Wang_Chan2018}
\bibinfo{author}{Wang, C.}, \bibinfo{author}{Chan, K.}, \bibinfo{year}{2018}.
\newblock \bibinfo{title}{Quasi-likelihood estimation of a censored
  autoregressive model with exogenous variables}.
\newblock \bibinfo{journal}{Journal of the American Statistical Association}
  \bibinfo{volume}{113:523}, \bibinfo{pages}{1135--1145}.
\bibitem[{Watanabe(2013)}]{Watanabe2013}
\bibinfo{author}{Watanabe, S.}, \bibinfo{year}{2013}.
\newblock \bibinfo{title}{A widely applicable bayesian information criterion}.
\newblock \bibinfo{journal}{Journal of Machine Learning Research}
  \bibinfo{volume}{14}, \bibinfo{pages}{867--897}.
\bibitem[{Wei and Tanner(1990)}]{Wei_1999}
\bibinfo{author}{Wei, G.C.G.}, \bibinfo{author}{Tanner, M.A.},
  \bibinfo{year}{1990}.
\newblock \bibinfo{title}{Posterior computations for censored regression data}.
\newblock \bibinfo{journal}{Journal of American Statistical Association}
  \bibinfo{volume}{85}, \bibinfo{pages}{829--839}.
\bibitem[{Zangari and Tsurumi(1996)}]{Zangari_Tsurumi1996}
\bibinfo{author}{Zangari, P.}, \bibinfo{author}{Tsurumi, H.},
  \bibinfo{year}{1996}.
\newblock \bibinfo{title}{A bayesian analysis of cansored autocorrelated data
  on exports of japanese pssenger cars to the united states}.
\newblock \bibinfo{journal}{Advances in Econometrics} \bibinfo{volume}{11, Part
  A}, \bibinfo{pages}{111--143}.
\bibitem[{Zeger and Brookmeyer(1986)}]{Zeger_Brookmeyer1984}
\bibinfo{author}{Zeger, S.}, \bibinfo{author}{Brookmeyer, R.},
  \bibinfo{year}{1986}.
\newblock \bibinfo{title}{Regression analysis with censored autocorrelated
  data}.
\newblock \bibinfo{journal}{Journal of the American Statistical Association}
  \bibinfo{volume}{81}, \bibinfo{pages}{722--729}.

\end{thebibliography}

\appendix
\section{Posterior Densities}
\label{sec:appA}

\begin{figure}[h]
  \includegraphics[scale=0.6]{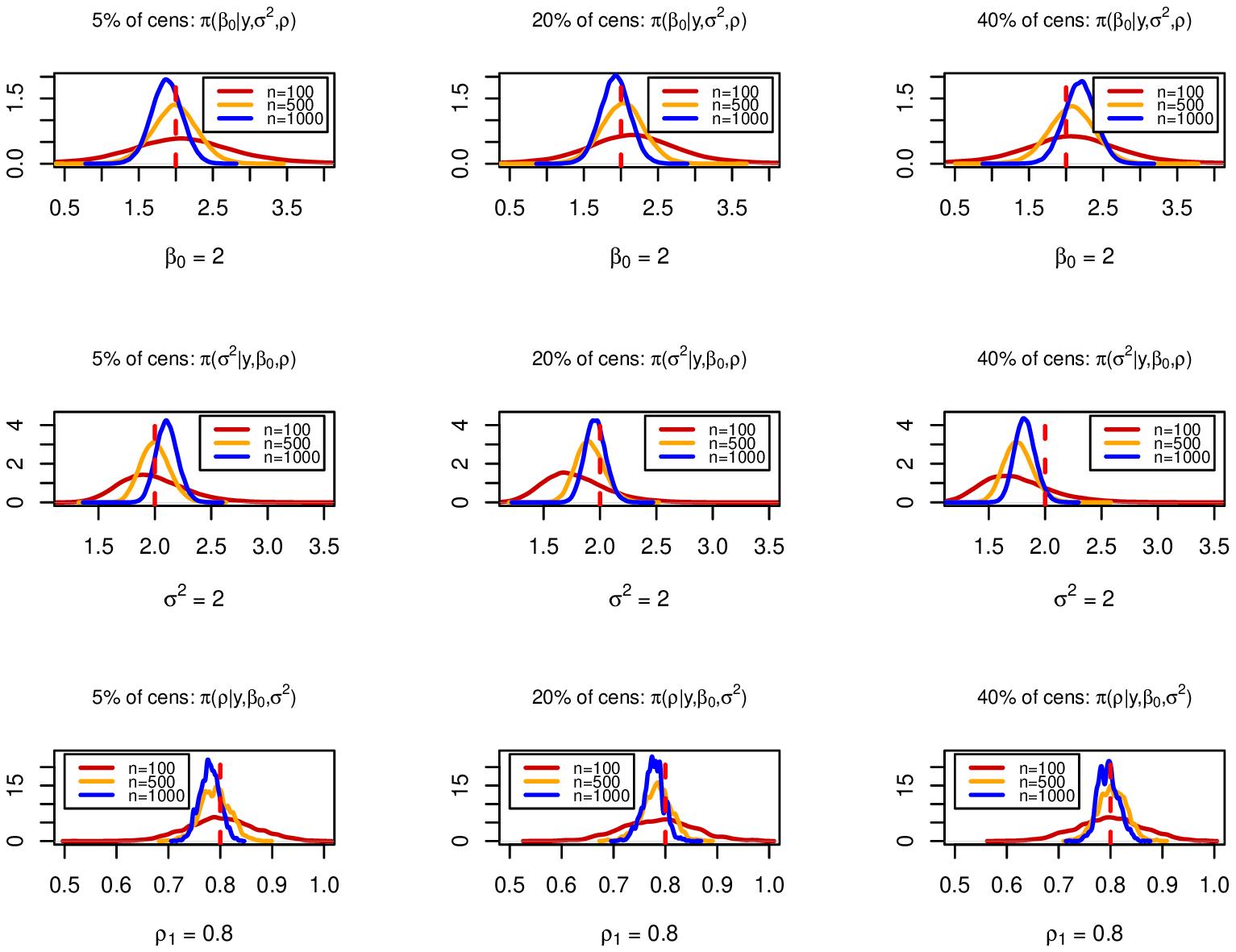}

  \includegraphics[scale=0.6]{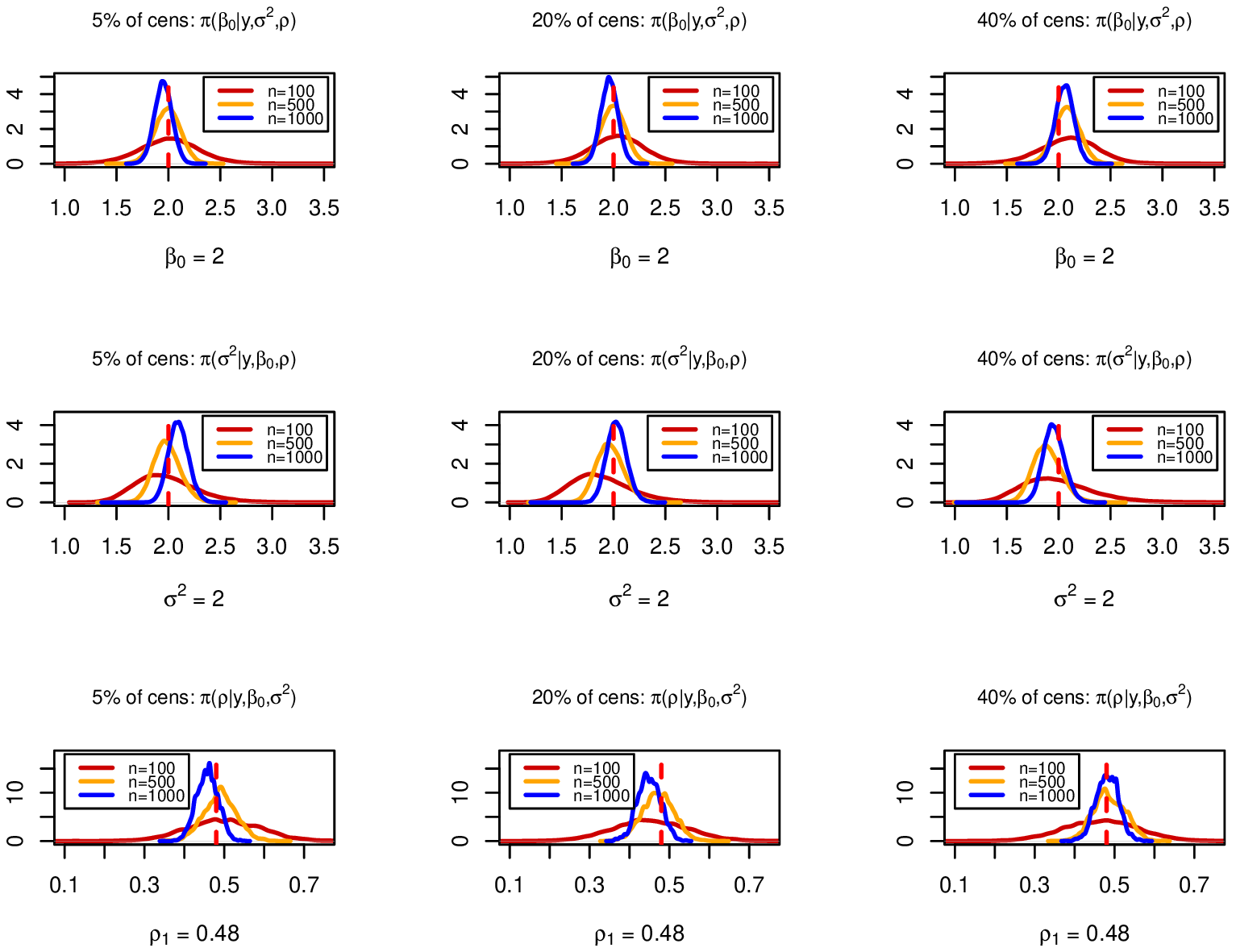}
        \caption{\textit{Model M1 with} $\rho=0.8$ top 3  lines and
          $\rho=0.48$ bottom 3 lines.  Posterior density of the model parameters for $n=100, 500$ and $1000$ under  3 censorship scenarios.}
	\label{fig:pd_M1_total} 
      \end{figure}

 \begin{figure}[h]
  \includegraphics[scale=0.5]{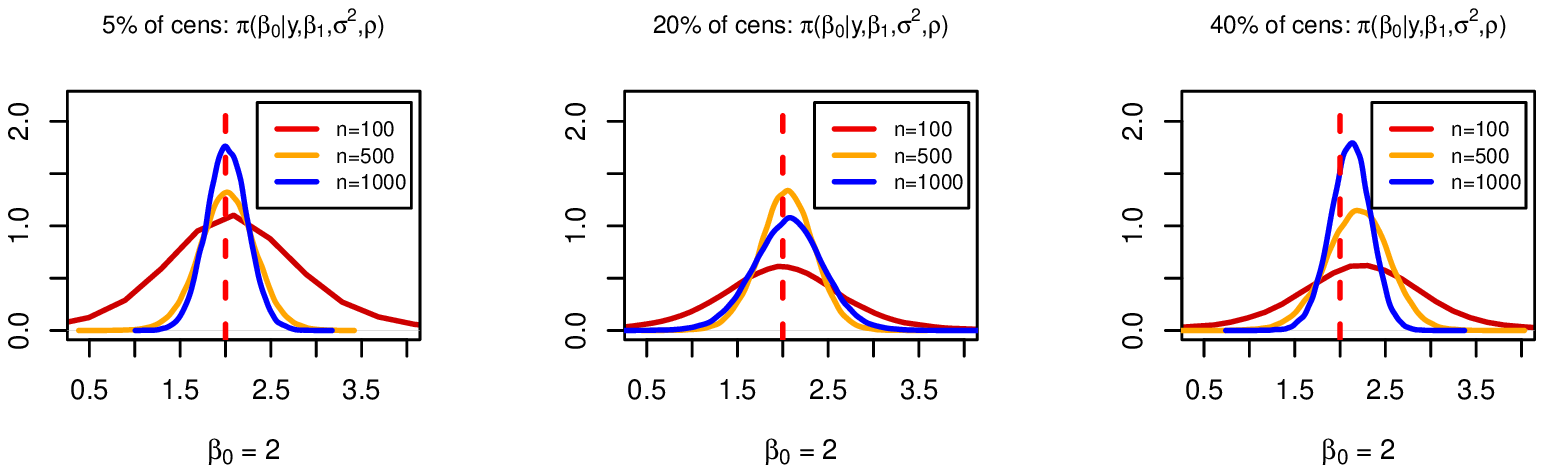}

  \includegraphics[scale=0.5]{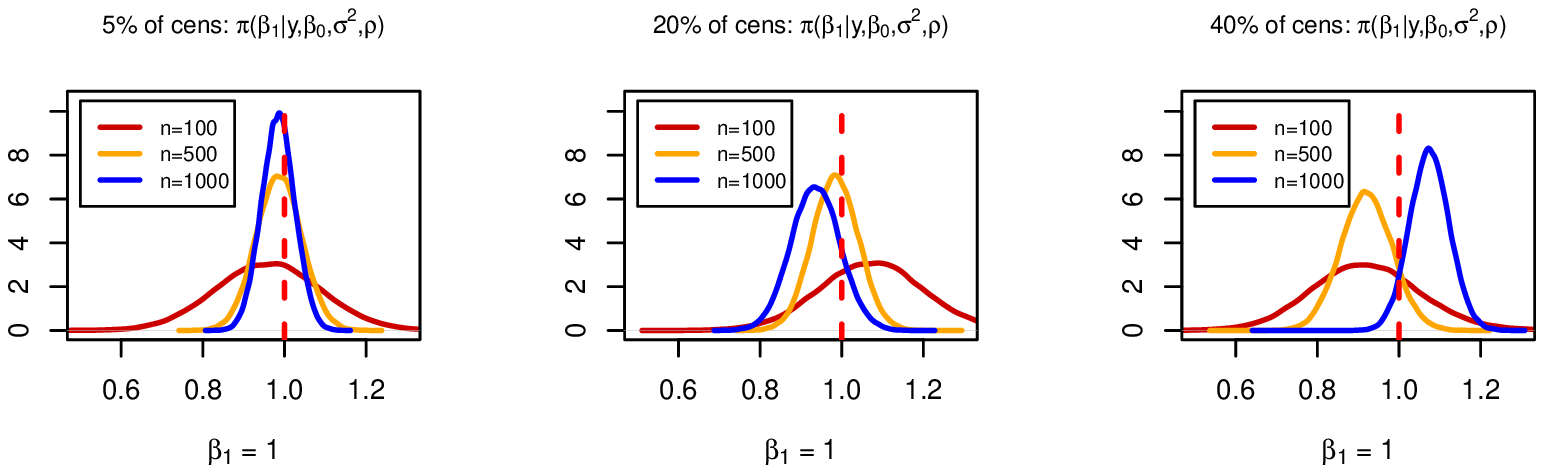}

  \includegraphics[scale=0.5]{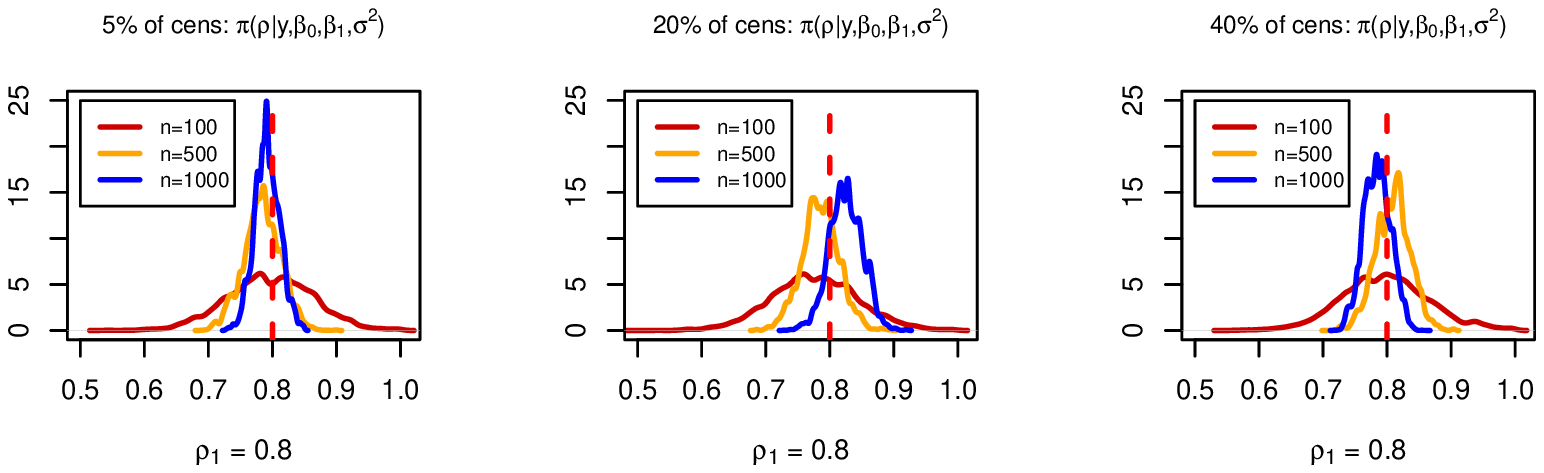}

  \includegraphics[scale=0.5]{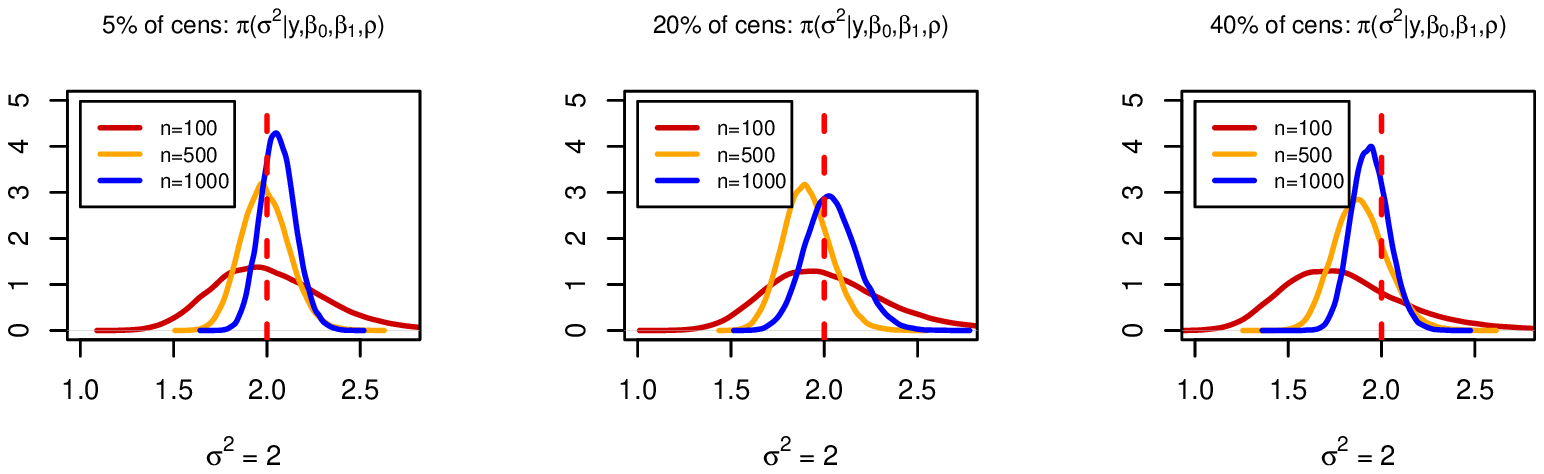}

	\caption{\textit{Model M2 with} $\rho=0.8$: Posterior density of the model parameters for $n=100, 500$ and $1000$ under  3 censorship scenarios.}
	\label{fig:pd_M2_1} 
      \end{figure}

 \begin{figure}[h]
  \includegraphics[scale=0.5]{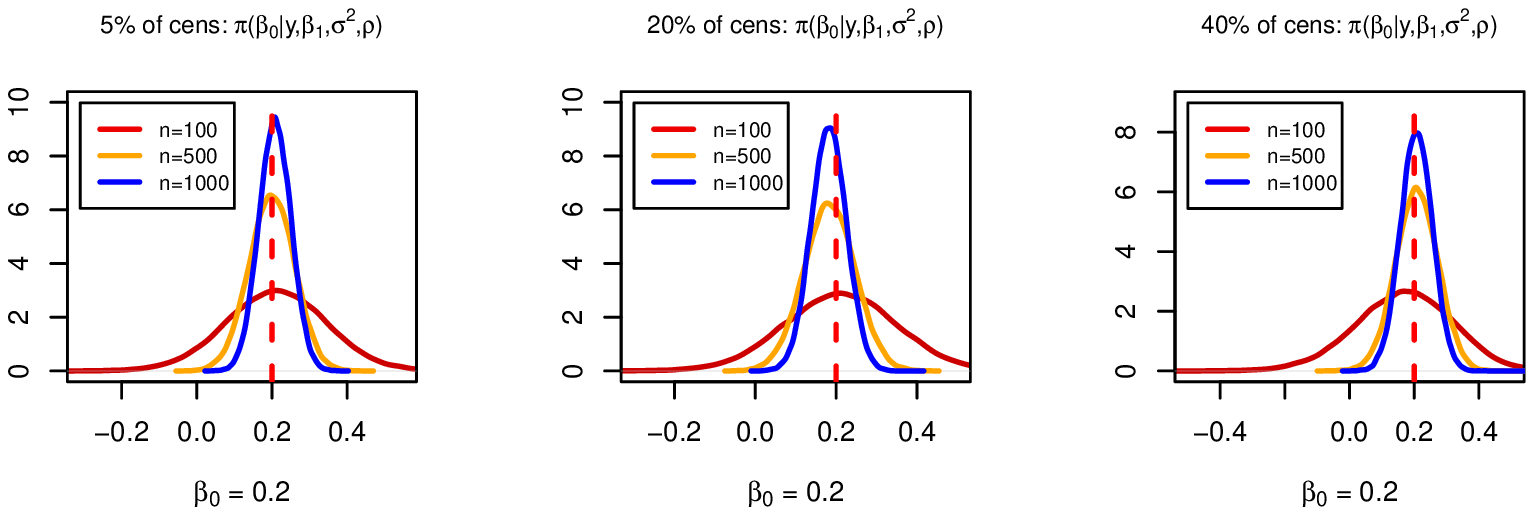}

  \includegraphics[scale=0.5]{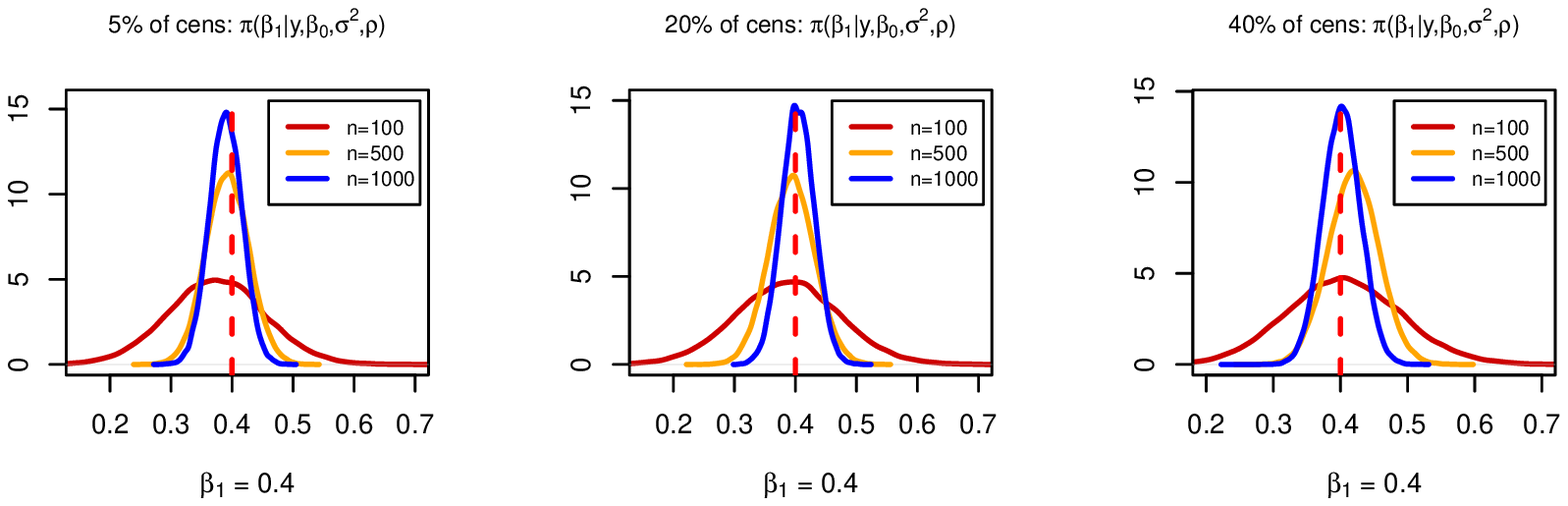}

  \includegraphics[scale=0.5]{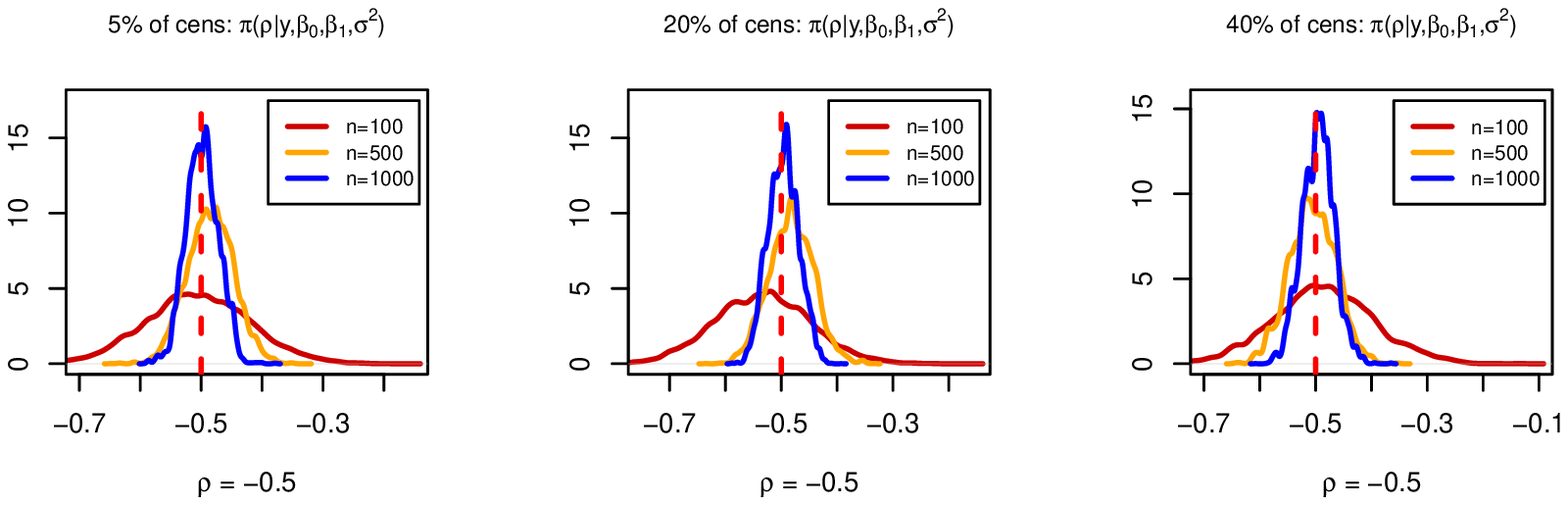}

  \includegraphics[scale=0.5]{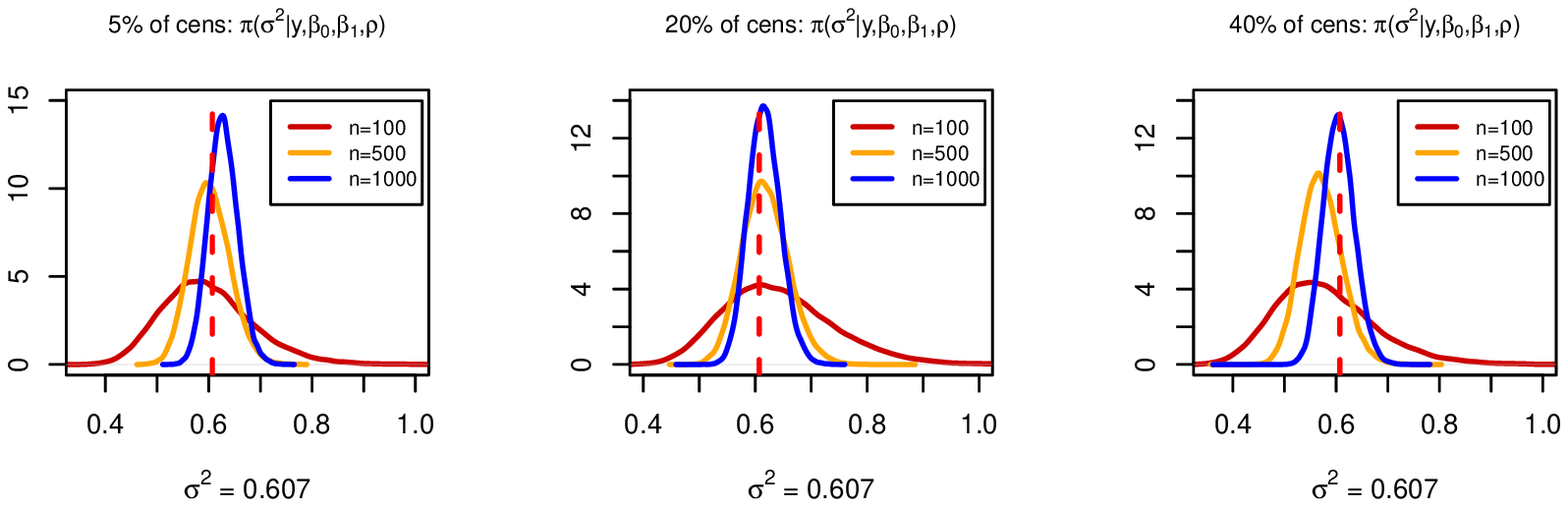}
  \caption{\textit{Model M3 with} $\rho=-0.5$: Posterior density of the model parameters for $n=100, 500$ and $1000$ under  3 censorship scenarios.}
	\label{fig:pd_M3_1} 
      \end{figure}

\clearpage
    

\section{Simulation Results}
\label{sec:appB}
\begin{table}[h!]
	\begin{center} 
		\caption{\textbf{Model 2}: Results, mean (standard
                  deviation), based on $100$ simulations of simple
                  CLR-AR($1$) model under different sample sizes and
                  censorship and $\bm{\rho>0}$.}
		\label{tab:model21} 
		\begin{tabular}{l l c c c c}
			\hline
			$n$ & $\%$ of cen &  $\beta_0=2$ & $\beta_1=1$ & $\sigma^2=2$ & $\rho=0.15$\\
			\hline 
			\multirow{3}{*}{$100$}&$5\%$&2.027{\scriptsize(0.066)}&0.995{\scriptsize(0.026)}&2.036{\scriptsize(0.107)}&0.134{\scriptsize(0.011)}\\
			&$20\%$&2.025{\scriptsize(0.070)}&0.995{\scriptsize(0.025)}&1.962{\scriptsize(0.107)}&0.133{\scriptsize(0.012)}\\
			&$40\%$&2.069{\scriptsize(0.100}&0.980{\scriptsize(0.030)}&1.756{\scriptsize(0.165)}&0.127{\scriptsize(0.016)}\\
			\hline 
			\multirow{3}{*}{$500$}&$5\%$&1.996{\scriptsize(0.018)}&1.006{\scriptsize(0.006)}&2.014{\scriptsize(0.017)}&0.148{\scriptsize(0.002)}\\
			&$20\%$&2.009{\scriptsize(0.018)}&1.002{\scriptsize(0.006)}&1.921{\scriptsize(0.025)}&0.147{\scriptsize(0.002)}\\
			&$40\%$&2.109{\scriptsize(0.026)}&0.968{\scriptsize(0.007)}&1.659{\scriptsize(0.135)}&0.140{\scriptsize(0.003)}\\
			\hline 
			\multirow{3}{*}{$1000$}&$5\%$&2.007{\scriptsize(0.009)}&0.995{\scriptsize(0.003)}&2.004{\scriptsize(0.009)}&0.149{\scriptsize(0.001)}\\
			&$20\%$&2.026{\scriptsize(0.011)}&0.987{\scriptsize(0.003)}&1.911{\scriptsize(0.017)}&0.145{\scriptsize(0.001)}\\
			&$40\%$&2.119{\scriptsize(0.025)}&0.956{\scriptsize(0.005)}&1.659{\scriptsize(0.125)}&0.141{\scriptsize(0.001)}\\
			\hhline{=|=|=|=|=|=}
			$n$ & $\%$ of cen &  $\beta_0=2$ & $\beta_1=1$ & $\sigma^2=2$ & $\rho=0.48$\\
			\hline 
			\multirow{3}{*}{$100$}&$5\%$&2.046{\scriptsize(0.085)}&0.992{\scriptsize(0.023)}&2.045{\scriptsize(0.106)}&0.460{\scriptsize(0.008)}\\
			&$20\%$&2.045{\scriptsize(0.098)}&0.992{\scriptsize(0.024)}&1.985{\scriptsize(0.120)}&0.452{\scriptsize(0.010)}\\
			&$40\%$&2.098{\scriptsize(0.119)}&0.983{\scriptsize(0.028)}&1.767{\scriptsize(0.179)}&0.450{\scriptsize(0.012)}\\
			\hline 
			\multirow{3}{*}{$500$}&$5\%$&2.001{\scriptsize(0.024)}&1.006{\scriptsize(0.006)}&2.015{\scriptsize(0.017)}&0.477{\scriptsize(0.002)}\\
			&$20\%$&2.020{\scriptsize(0.027)}&1.000{\scriptsize(0.007)}&1.929{\scriptsize(0.023)}&0.472{\scriptsize(0.002)}\\
			&$40\%$&2.125{\scriptsize(0.038)}&0.966{\scriptsize(0.008)}&1.696{\scriptsize(0.115)}&0.463{\scriptsize(0.002)}\\
			\hline 
			\multirow{3}{*}{$1000$}&$5\%$&2.009{\scriptsize(0.013)}&0.995{\scriptsize(0.003)}&2.004{\scriptsize(0.009)}&0.477{\scriptsize(0.001)}\\
			&$20\%$&2.031{\scriptsize(0.016)}&0.987{\scriptsize(0.003)}&1.924{\scriptsize(0.016)}&0.471{\scriptsize(0.001)}\\
			&$40\%$&2.130{\scriptsize(0.032)}&0.957{\scriptsize(0.006)}&1.689{\scriptsize(0.108)}&0.462{\scriptsize(0.001)}\\
			\hhline{=|=|=|=|=|=}
			$n$ & $\%$ of cen &  $\beta_0=2$ & $\beta_1=1$ & $\sigma^2=2$ & $\rho=0.8$\\
			\hline 
			\multirow{3}{*}{$100$}&$5\%$&2.110{\scriptsize(0.354)}&0.992{\scriptsize(0.019)}&2.106 {\scriptsize(0.105)}&0.780{\scriptsize(0.004)}\\
			&$20\%$&2.130{\scriptsize(0.369)}&0.999{\scriptsize(0.019)}&1.991{\scriptsize(0.111)}&0.777{\scriptsize(0.004)}\\
			&$40\%$&2.290{\scriptsize(0.397)}&0.968{\scriptsize(0.027)}&1.755{\scriptsize(0.214)}&0.762{\scriptsize(0.007)}\\
			\hline 
			\multirow{3}{*}{$500$}&$5\%$&2.022{\scriptsize(0.099)}&1.005{\scriptsize(0.005)}&2.016{\scriptsize(0.017)}&0.793{\scriptsize(0.001)}\\
			&$20\%$&2.054{\scriptsize(0.101)}&1.000{\scriptsize(0.005)}&1.942{\scriptsize(0.022)}&0.789{\scriptsize(0.001)}\\
			&$40\%$&2.200{\scriptsize(0.119)}&0.979{\scriptsize(0.007)}&1.714{\scriptsize(0.107)}&0.779{\scriptsize(0.002)}\\
			\hline 
			\multirow{3}{*}{$1000$}&$5\%$&2.013{\scriptsize(0.058)}&0.995{\scriptsize(0.002)}&2.005{\scriptsize(0.009)}&0.795{\scriptsize(0.000)}\\
			&$20\%$&2.054{\scriptsize(0.101)}&1.000{\scriptsize(0.005)}&1.942{\scriptsize(0.022)}&0.789{\scriptsize(0.001)}\\
			&$40\%$&2.200{\scriptsize(0.119)}&0.979{\scriptsize(0.007)}&1.714{\scriptsize(0.107)}&0.779{\scriptsize(0.002)}\\
			\hline
		\end{tabular}
	\end{center}
\end{table}
%
\begin{table}[h!]
	\begin{center} 
		\caption{\textbf{Model 2}: Results based on $100$
                  simulations of simple CLR-AR($1$) model under
                  different sample sizes and censorship, using GDA
                  with mean of multiple samples and $\bm{\rho<0}$.} 
		\label{tab:model22} 
		\begin{tabular}{l l c c c c}
			\hline
			$n$ & $\%$ of cen &  $\beta_0=2$ & $\beta_1=1$ & $\sigma^2=2$ & $\rho=-0.15$\\
			\hline 
			\multirow{3}{*}{$100$}&$5\%$&2.013{\scriptsize(0.056)}&1.000{\scriptsize(0.023)}&2.041{\scriptsize(0.109)}&-0.161{\scriptsize(0.009)}\\
			&$20\%$&2.006{\scriptsize(0.060)}&1.003{\scriptsize(0.023)}&1.960{\scriptsize(0.098)}&-0.160{\scriptsize(0.012)}\\
			&$40\%$&2.066{\scriptsize(0.076}&0.980{\scriptsize(0.024)}&1.760{\scriptsize(0.162)}&-0.160{\scriptsize(0.014)}\\
			\hline 
			\multirow{3}{*}{$500$}&$5\%$&1.994{\scriptsize(0.015)}&1.007{\scriptsize(0.006)}&2.014{\scriptsize(0.017)}&-0.152{\scriptsize(0.002)}\\
			&$20\%$&2.011{\scriptsize(0.016)}&1.000{\scriptsize(0.006)}&1.914{\scriptsize(0.026)}&-0.151{\scriptsize(0.002)}\\
			&$40\%$&2.090{\scriptsize(0.023)}&0.974{\scriptsize(0.006)}&1.672{\scriptsize(0.125)}&-0.151{\scriptsize(0.003)}\\
			\hline 
			\multirow{3}{*}{$1000$}&$5\%$&2.005{\scriptsize(0.007)}&0.997{\scriptsize(0.002)}&2.005{\scriptsize(0.009)}&-0.151{\scriptsize(0.001)}\\
			&$20\%$&2.011{\scriptsize(0.016)}&1.000{\scriptsize(0.006)}&1.917{\scriptsize(0.025)}&-0.151{\scriptsize(0.002)}\\
			&$40\%$&2.119{\scriptsize(0.025)}&0.956{\scriptsize(0.005)}&1.655{\scriptsize(0.128)}&-0.148{\scriptsize(0.001)}\\
			\hhline{=|=|=|=|=|=} 
			$n$ & $\%$ of cen &  $\beta_0=2$ & $\beta_1=1$ & $\sigma^2=2$ & $\rho=-0.48$\\
			\hline 
			\multirow{3}{*}{$100$}&$5\%$&2.003{\scriptsize(0.045)}&1.004{\scriptsize(0.017)}&2.055{\scriptsize(0.115)}&-0.481{\scriptsize(0.007)}\\
			&$20\%$&1.992{\scriptsize(0.057)}&1.009{\scriptsize(0.018)}&1.989{\scriptsize(0.102)}&-0.479{\scriptsize(0.008)}\\
			&$40\%$&2.054{\scriptsize(0.065)}&0.985{\scriptsize(0.018)}&1.837{\scriptsize(0.123)}&-0.467{\scriptsize(0.009)}\\
			\hline 
			\multirow{3}{*}{$500$}&$5\%$&1.992{\scriptsize(0.011)}&1.007{\scriptsize(0.004)}&2.021{\scriptsize(0.017)}&-0.479{\scriptsize(0.002)}\\
			&$20\%$&2.003{\scriptsize(0.011)}&1.003{\scriptsize(0.004)}&1.919{\scriptsize(0.030)}&-0.475{\scriptsize(0.002)}\\
			&$40\%$&2.090{\scriptsize(0.023)}&0.972{\scriptsize(0.005)}&1.741{\scriptsize(0.086)}&-0.474{\scriptsize(0.002)}\\
			\hline 
			\multirow{3}{*}{$1000$}&$5\%$&2.001{\scriptsize(0.005)}&0.999{\scriptsize(0.002)}&2.011{\scriptsize(0.009)}&-0.479{\scriptsize(0.001)}\\
			&$20\%$&2.012{\scriptsize(0.006)}&0.995{\scriptsize(0.002)}&1.940{\scriptsize(0.014)}&-0.476{\scriptsize(0.001)}\\
			&$40\%$&2.097{\scriptsize(0.018)}&0.965{\scriptsize(0.004)}&1.740{\scriptsize(0.078)}&-0.468{\scriptsize(0.001)}\\
			\hhline{=|=|=|=|=|=}
			$n$ & $\%$ of cen &  $\beta_0=2$ & $\beta_1=1$ & $\sigma^2=2$ & $\rho=-0.8$\\
			\hline 
			\multirow{3}{*}{$100$}&$5\%$&1.992{\scriptsize(0.033)}&1.007{\scriptsize(0.012)}&2.089 {\scriptsize(0.120)}&-0.797{\scriptsize(0.003)}\\
			&$20\%$&1.994{\scriptsize(0.052)}&1.003{\scriptsize(0.015)}&2.097{\scriptsize(0.134)}&-0.793{\scriptsize(0.003)}\\
			&$40\%$&2.065{\scriptsize(0.103)}&0.969{\scriptsize(0.029)}&2.059{\scriptsize(0.141)}&-0.780{\scriptsize(0.005)}\\
			\hline 
			\multirow{3}{*}{$500$}&$5\%$&1.991{\scriptsize(0.010)}&1.007{\scriptsize(0.003)}&2.033{\scriptsize(0.018)}&-0.793{\scriptsize(0.001)}\\
			&$20\%$&1.998{\scriptsize(0.010)}&1.007{\scriptsize(0.003)}&2.025{\scriptsize(0.021)}&-0.794{\scriptsize(0.001)}\\
			&$40\%$&2.071{\scriptsize(0.023)}&0.977{\scriptsize(0.006)}&1.947{\scriptsize(0.030)}&-0.786{\scriptsize(0.001)}\\
			\hline 
			\multirow{3}{*}{$1000$}&$5\%$&1.999{\scriptsize(0.004)}&1.000{\scriptsize(0.001)}&2.019{\scriptsize(0.009)}&-0.797{\scriptsize(0.000)}\\
			&$20\%$&1.992{\scriptsize(0.010)}&1.006{\scriptsize(0.003)}&2.025{\scriptsize(0.021)}&-0.794{\scriptsize(0.001)}\\
			&$40\%$&2.061{\scriptsize(0.013)}&0.977{\scriptsize(0.003)}&1.959{\scriptsize(0.016)}&-0.784{\scriptsize(0.001)}\\
			\hline
		\end{tabular}
	\end{center}
\end{table}
%
%
%
\begin{table}[h!]
	\begin{center} 
		\caption{\textbf{Model 1}: Results based on $100$
                  simulations of CLR-AR($1$) model without explanatory
                  variables, under different sample sizes and
                  censorship, using GDA with mean of multiple samples and $\bm{\rho>0}$.}
		\label{tab:model11} 
		\begin{tabular}{l l c c c}
			\hline
			$n$ & $\%$ of cen &  $\beta_0=2$ & $\sigma^2=2$ & $\rho=0.15$\\
			\hline 
			\multirow{3}{*}{$100$}&$5\%$&2.021{\scriptsize(0.021)}&2.037{\scriptsize(0.108)}&0.133{\scriptsize(0.010)}\\
			&$20\%$&2.021{\scriptsize(0.021)}&1.974{\scriptsize(0.111)}&0.131{\scriptsize(0.011)}\\
			&$40\%$&2.066{\scriptsize(0.023)}&1.693{\scriptsize(0.208)}&0.125{\scriptsize(0.014)}\\
			\hline 
			\multirow{3}{*}{$500$}&$5\%$&2.006{\scriptsize(0.058)}&2.015{\scriptsize(0.017)}&0.149{\scriptsize(0.002)}\\
			&$20\%$&2.011{\scriptsize(0.006)}&1.931{\scriptsize(0.023)}&0.147{\scriptsize(0.002)}\\
			&$40\%$&2.053{\scriptsize(0.008)}&1.681{\scriptsize(0.120)}&0.142{\scriptsize(0.003)}\\
			\hline 
			\multirow{3}{*}{$1000$}&$5\%$&2.000{\scriptsize(0.003)}&2.005{\scriptsize(0.009)}&0.148{\scriptsize(0.001)}\\
			&$20\%$&2.005{\scriptsize(0.003)}&1.923{\scriptsize(0.016)}&0.148{\scriptsize(0.001)}\\
			&$40\%$&2.046{\scriptsize(0.004)}&1.673{\scriptsize(0.116)}&0.143{\scriptsize(0.001)}\\
			\hhline{=|=|=|=|=}
			$n$ & $\%$ of cen &  $\beta_0=2$ & $\sigma^2=2$ & $\rho=0.48$\\
			\hline 
			\multirow{3}{*}{$100$}&$5\%$&2.034{\scriptsize(0.055)}&2.044{\scriptsize(0.105)}&0.461{\scriptsize(0.008)}\\
			&$20\%$&2.037{\scriptsize(0.055)}&1.993{\scriptsize(0.135)}&0.456{\scriptsize(0.008)}\\
			&$40\%$&2.087{\scriptsize(0.056)}&1.753{\scriptsize(0.195)}&0.444{\scriptsize(0.012)}\\
			\hline 
			\multirow{3}{*}{$500$}&$5\%$&2.010{\scriptsize(0.015)}&2.015{\scriptsize(0.016)}&0.477{\scriptsize(0.002)}\\
			&$20\%$&2.018{\scriptsize(0.015)}&1.946{\scriptsize(0.023)}&0.473{\scriptsize(0.002)}\\
			&$40\%$&2.067{\scriptsize(0.017)}&1.725{\scriptsize(0.096)}&0.462{\scriptsize(0.002)}\\
			\hline 
			\multirow{3}{*}{$1000$}&$5\%$&2.001{\scriptsize(0.008)}&2.001{\scriptsize(0.009)}&0.477{\scriptsize(0.001)}\\
			&$20\%$&2.008{\scriptsize(0.008)}&1.938{\scriptsize(0.014)}&0.472{\scriptsize(0.001)}\\
			&$40\%$&2.058{\scriptsize(0.010)}&1.717{\scriptsize(0.090)}&0.461{\scriptsize(0.001)}\\
			\hhline{=|=|=|=|=}
			$n$ & $\%$ of cen &  $\beta_0=2$ & $\sigma^2=2$ & $\rho=0.8$\\
			\hline 
			\multirow{3}{*}{$100$}&$5\%$&2.097{\scriptsize(0.368)}&2.056{\scriptsize(0.111)}&0.781{\scriptsize(0.004)}\\
			&$20\%$&2.126{\scriptsize(0.346)}&2.000{\scriptsize(0.122)}&0.774{\scriptsize(0.004)}\\
			&$40\%$&2.266{\scriptsize(0.341)}&1.708{\scriptsize(0.231)}&0.762{\scriptsize(0.006)}\\
			\hline 
			\multirow{3}{*}{$500$}&$5\%$&2.027{\scriptsize(0.097)}&2.018{\scriptsize(0.019)}&0.793{\scriptsize(0.001)}\\
			&$20\%$&2.054{\scriptsize(0.095)}&1.939{\scriptsize(0.021)}&0.788{\scriptsize(0.001)}\\
			&$40\%$&2.191{\scriptsize(0.109)}&1.673{\scriptsize(0.133)}&0.778{\scriptsize(0.002)}\\
			\hline 
			\multirow{3}{*}{$1000$}&$5\%$&2.003{\scriptsize(0.054)}&2.009{\scriptsize(0.009)}&0.795{\scriptsize(0.000)}\\
			&$20\%$&2.036{\scriptsize(0.053)}&1.926{\scriptsize(0.016)}&0.789{\scriptsize(0.001)}\\
			&$40\%$&2.157{\scriptsize(0.065)}&1.682{\scriptsize(0.115)}&0.780{\scriptsize(0.001)}\\
			\hline
		\end{tabular}
	\end{center}
\end{table}
%
\begin{table}[h!]
	\begin{center} 
		\caption{\textbf{Model 1}: Results based on $100$ simulations of CLR-AR($1$) model without explanatory variables, under different sample sizes and censorship, using GDA with mean of multiple samples and $\bm{\rho<0}$.}
		\label{tab:model12} 
		\begin{tabular}{l l c c c}
			\hline
			$n$ & $\%$ of cen &  $\beta_0=2$ & $\sigma^2=2$ & $\rho=-0.15$\\
			\hline 
			\multirow{3}{*}{$100$}&$5\%$&2.017{\scriptsize(0.012)}&2.035{\scriptsize(0.107)}&-0.163{\scriptsize(0.010)}\\
			&$20\%$&2.011{\scriptsize(0.013)}&1.972{\scriptsize(0.104)}&-0.162{\scriptsize(0.011)}\\
			&$40\%$&2.051{\scriptsize(0.015)}&1.721{\scriptsize(0.174)}&-0.169{\scriptsize(0.013)}\\
			\hline 
			\multirow{3}{*}{$500$}&$5\%$&2.004{\scriptsize(0.003)}&2.017{\scriptsize(0.018)}&-0.152{\scriptsize(0.002)}\\
			&$20\%$&2.008{\scriptsize(0.003)}&1.932{\scriptsize(0.023)}&-0.152{\scriptsize(0.002)}\\
			&$40\%$&2.046{\scriptsize(0.005)}&1.688{\scriptsize(0.115)}&-0.154{\scriptsize(0.003)}\\
			\hline 
			\multirow{3}{*}{$1000$}&$5\%$&2.000{\scriptsize(0.002)}&2.004{\scriptsize(0.009)}&-0.151{\scriptsize(0.001)}\\
			&$20\%$&2.004{\scriptsize(0.002)}&1.925{\scriptsize(0.015)}&-0.150{\scriptsize(0.001)}\\
			&$40\%$&2.045{\scriptsize(0.004)}&1.674{\scriptsize(0.115)}&-0.152{\scriptsize(0.001)}\\
			\hhline{=|=|=|=|=}
			$n$ & $\%$ of cen &  $\beta_0=2$ & $\sigma^2=2$ & $\rho=-0.48$\\
			\hline 
			\multirow{3}{*}{$100$}&$5\%$&2.013{\scriptsize(0.007)}&2.048{\scriptsize(0.112)}&-0.487{\scriptsize(0.007)}\\
			&$20\%$&2.008{\scriptsize(0.009)}&1.996{\scriptsize(0.114)}&-0.486{\scriptsize(0.008)}\\
			&$40\%$&2.026{\scriptsize(0.013)}&1.850{\scriptsize(0.139)}&-0.478{\scriptsize(0.008)}\\
			\hline 
			\multirow{3}{*}{$500$}&$5\%$&2.003{\scriptsize(0.002)}&2.021{\scriptsize(0.018)}&-0.480{\scriptsize(0.002)}\\
			&$20\%$&2.007{\scriptsize(0.003)}&1.954{\scriptsize(0.021)}&-0.478{\scriptsize(0.002)}\\
			&$40\%$&2.047{\scriptsize(0.005)}&1.714{\scriptsize(0.100)}&-0.481{\scriptsize(0.002)}\\
			\hline 
			\multirow{3}{*}{$1000$}&$5\%$&2.000{\scriptsize(0.001)}&2.008{\scriptsize(0.009)}&-0.479{\scriptsize(0.001)}\\
			&$20\%$&2.004{\scriptsize(0.001)}&1.947{\scriptsize(0.013)}&-0.476{\scriptsize(0.001)}\\
			&$40\%$&2.040{\scriptsize(0.003)}&1.755{\scriptsize(0.070)}&-0.471{\scriptsize(0.001)}\\
			\hhline{=|=|=|=|=}
			$n$ & $\%$ of cen &  $\beta_0=2$ & $\sigma^2=2$ & $\rho=-0.8$\\
			\hline 
			\multirow{3}{*}{$100$}&$5\%$&2.002{\scriptsize(0.005)}&2.068{\scriptsize(0.108)}&-0.790{\scriptsize(0.003)}\\
			&$20\%$&1.993{\scriptsize(0.007)}&2.051{\scriptsize(0.118)}&-0.787{\scriptsize(0.004)}\\
			&$40\%$&2.012{\scriptsize(0.013)}&1.953{\scriptsize(0.122)}&-0.783{\scriptsize(0.004)}\\
			\hline 
			\multirow{3}{*}{$500$}&$5\%$&2.000{\scriptsize(0.001)}&2.029{\scriptsize(0.019)}&-0.797{\scriptsize(0.001)}\\
			&$20\%$&2.003{\scriptsize(0.002)}&1.984{\scriptsize(0.020)}&0.796{\scriptsize(0.001)}\\
			&$40\%$&2.029{\scriptsize(0.004)}&1.882{\scriptsize(0.039)}&-0.793{\scriptsize(0.001)}\\
			\hline 
			\multirow{3}{*}{$1000$}&$5\%$&1.999{\scriptsize(0.001)}&2.015{\scriptsize(0.010)}&-0.797{\scriptsize(0.000)}\\
			&$20\%$&2.001{\scriptsize(0.001)}&1.982{\scriptsize(0.011)}&-0.795{\scriptsize(0.000)}\\
			&$40\%$&2.025{\scriptsize(0.002)}&1.885{\scriptsize(0.025)}&-0.791{\scriptsize(0.000)}\\
			\hline
		\end{tabular}
	\end{center}
\end{table}

      \clearpage

\section{Measures of predictive performance} \label{sec:appwaic}

\textit{DIC}\\

  DIC is a measure of fit quality widely used in Bayesian approach and is calculated as follows:
\begin{equation}\label{eq:dic}
	DIC=\displaystyle{-4\cdot E_{\theta\lvert\mathbf{y}}[ln f(\mathbf{y}\lvert\theta)]+2\cdot ln f(\mathbf{y}\lvert\hat{\theta})},
\end{equation}
where $E_{\theta\lvert\mathbf{y}}[ln f(\mathbf{y}\lvert\theta)]$ is the posterior mean of the log-likelihood function, given by
\begin{equation}\label{eq:lk_pm}
	E_{\theta\lvert\mathbf{y}}[ln f(\mathbf{y}\lvert\theta)]=\displaystyle{\frac{1}{M}\sum_{j=1}^{M}ln f(\mathbf{y}\lvert\bm{\theta}^{(j)})},
\end{equation}
and $f(\mathbf{y}\lvert\hat{\theta})$ is the likelihood function evaluated at the Bayesian parameters estimates.\\

 \textit{WAIC}\\

  WAIC is another measure of predictive accuracy, more related to Bayesian approach than previous criterion \cite{Watanabe2013}, and is given by\\
\begin{equation}\label{eq:waic}
	WAIC=\displaystyle{-2\sum_{t=T}^{n} ln E_{\theta\lvert\mathbf{y}} [f(y_t\lvert\theta)]+2pw}.
\end{equation}
where $pw$ is the correction term, often as the one used in DIC criterion, defined as follows:
\begin{equation}\label{eq:pwa}
	pw=\displaystyle{-2\sum_{t=1}^{T} \{E_{\theta\lvert\mathbf{y}} [ln f(y_t\lvert\theta)]-ln E_{\theta\lvert\mathbf{y}}[f(y_t\lvert\theta)]\}}.
\end{equation}

\clearpage

\section{Analysis of the convergence of the
  chains} \label{sec:convergence}

\begin{figure}[t] 
		\includegraphics[width=6cm]{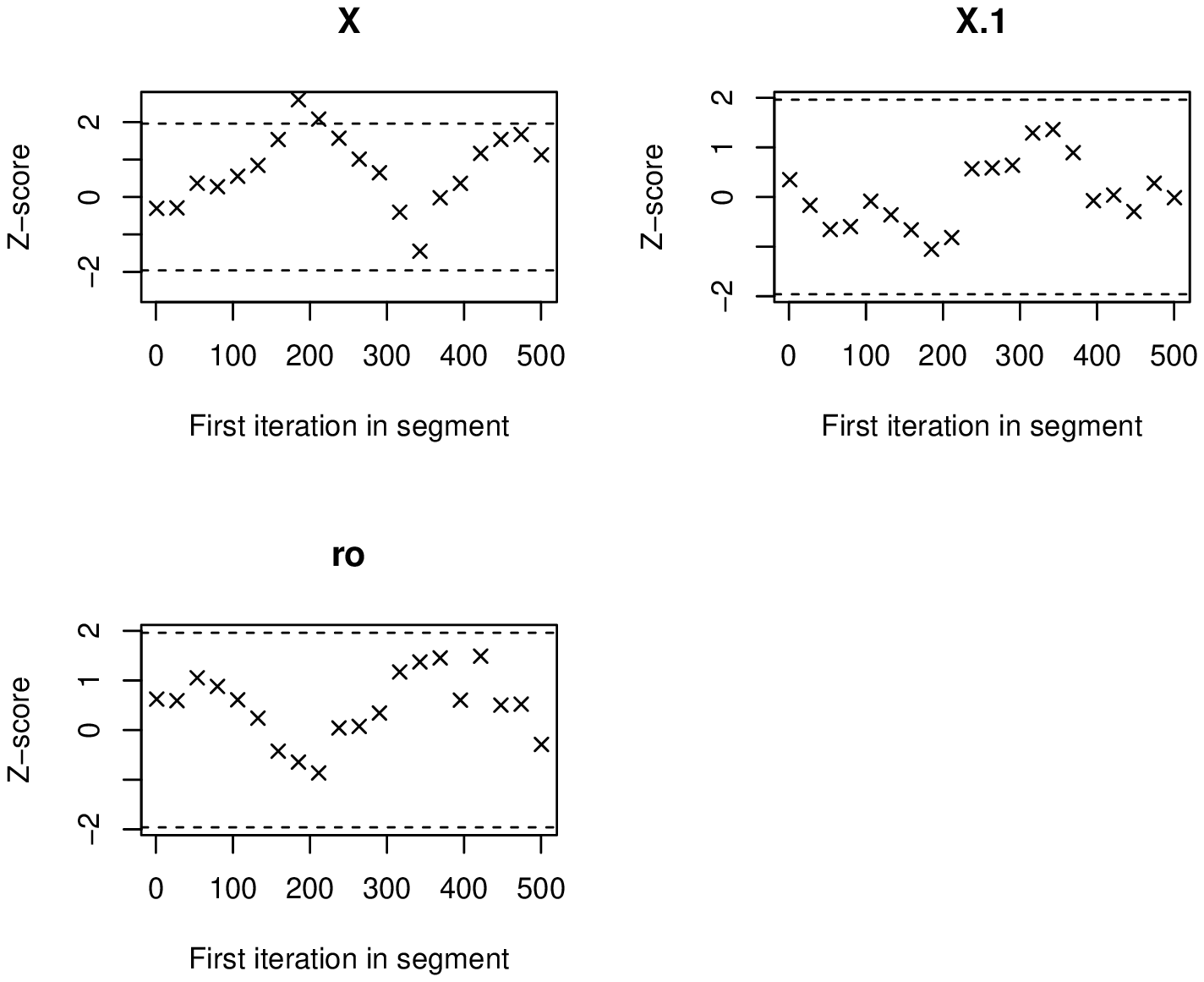} 
		\includegraphics[width=6cm]{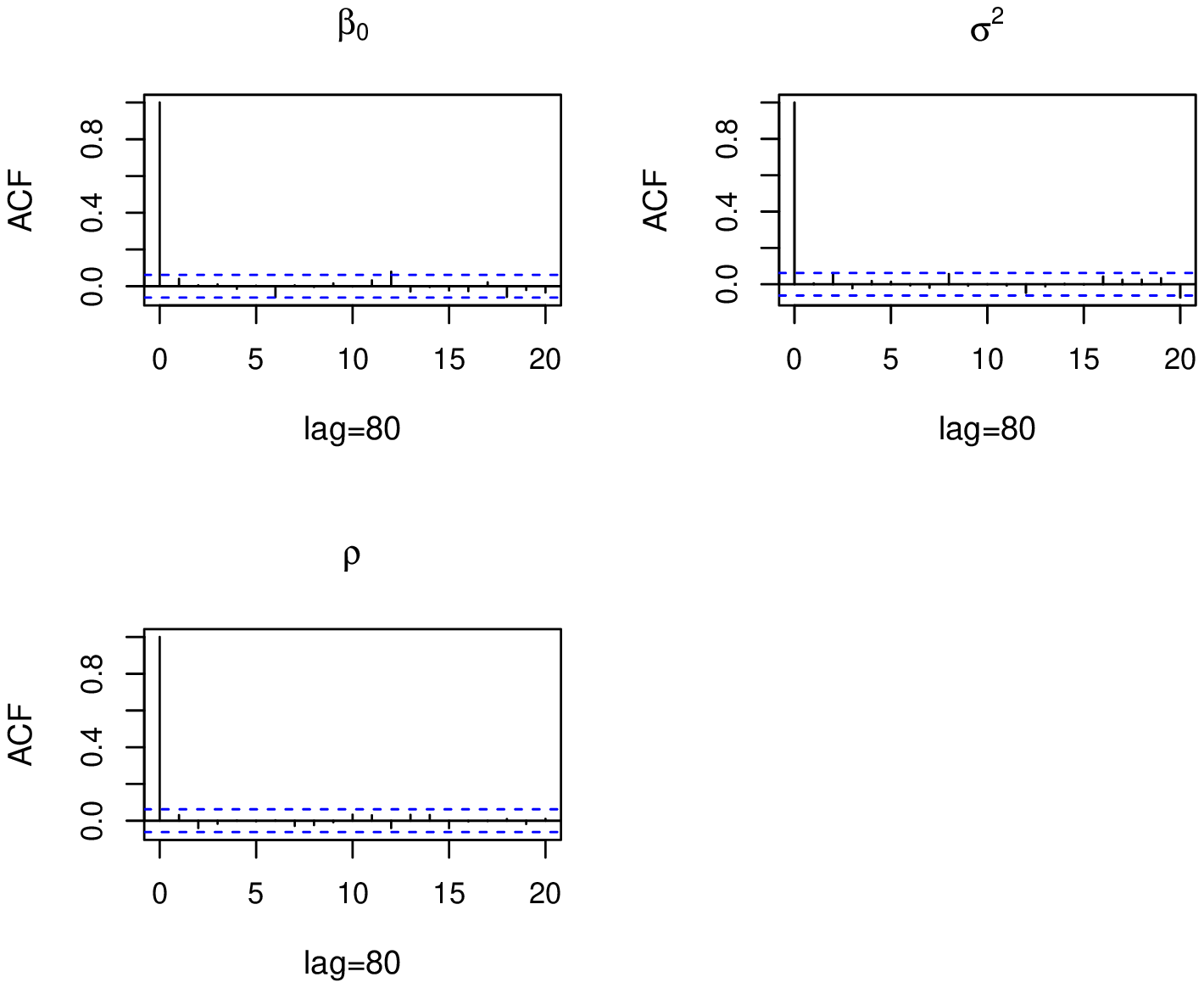} 
		\includegraphics[width=6cm]{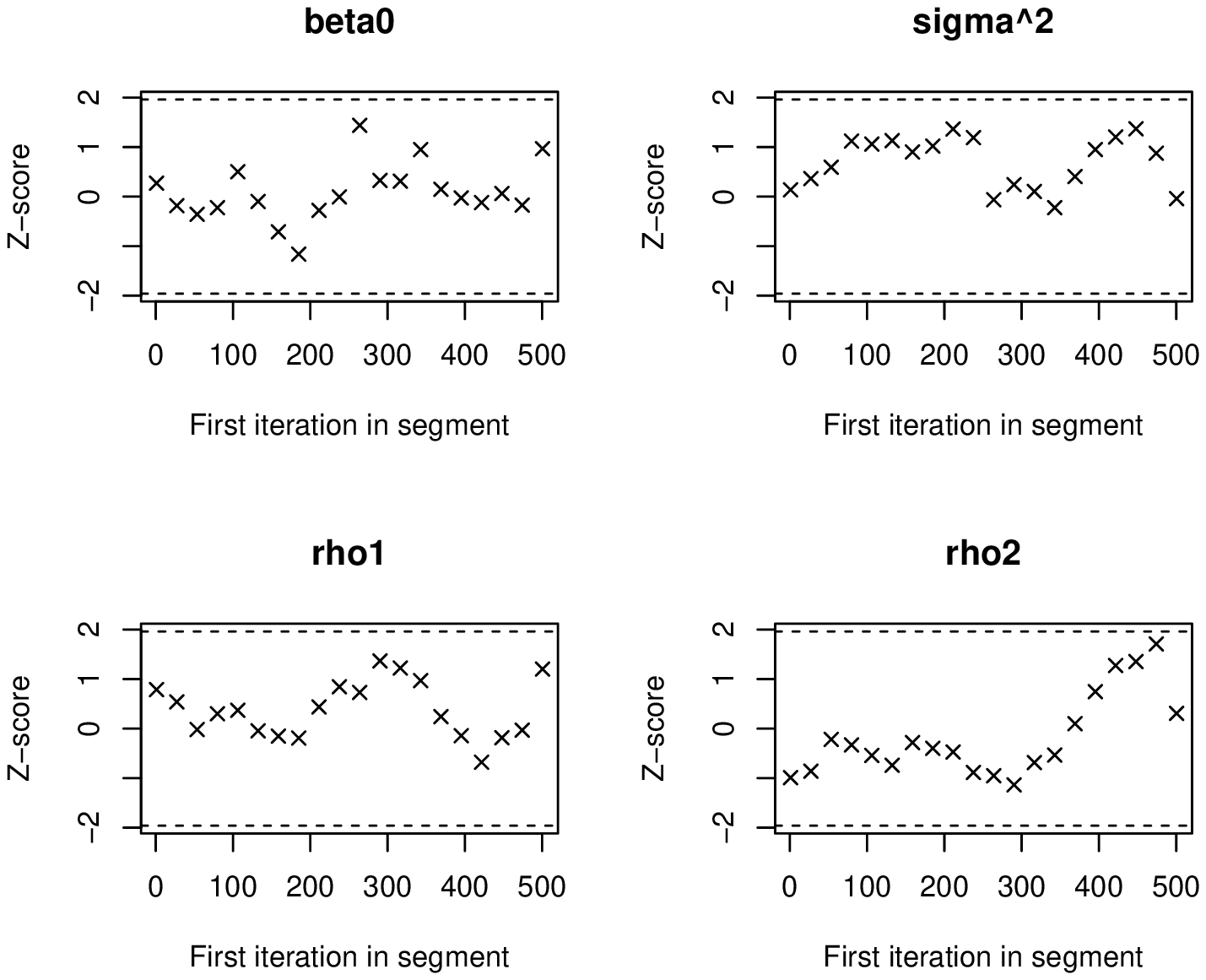}   
		\includegraphics[width=6cm]{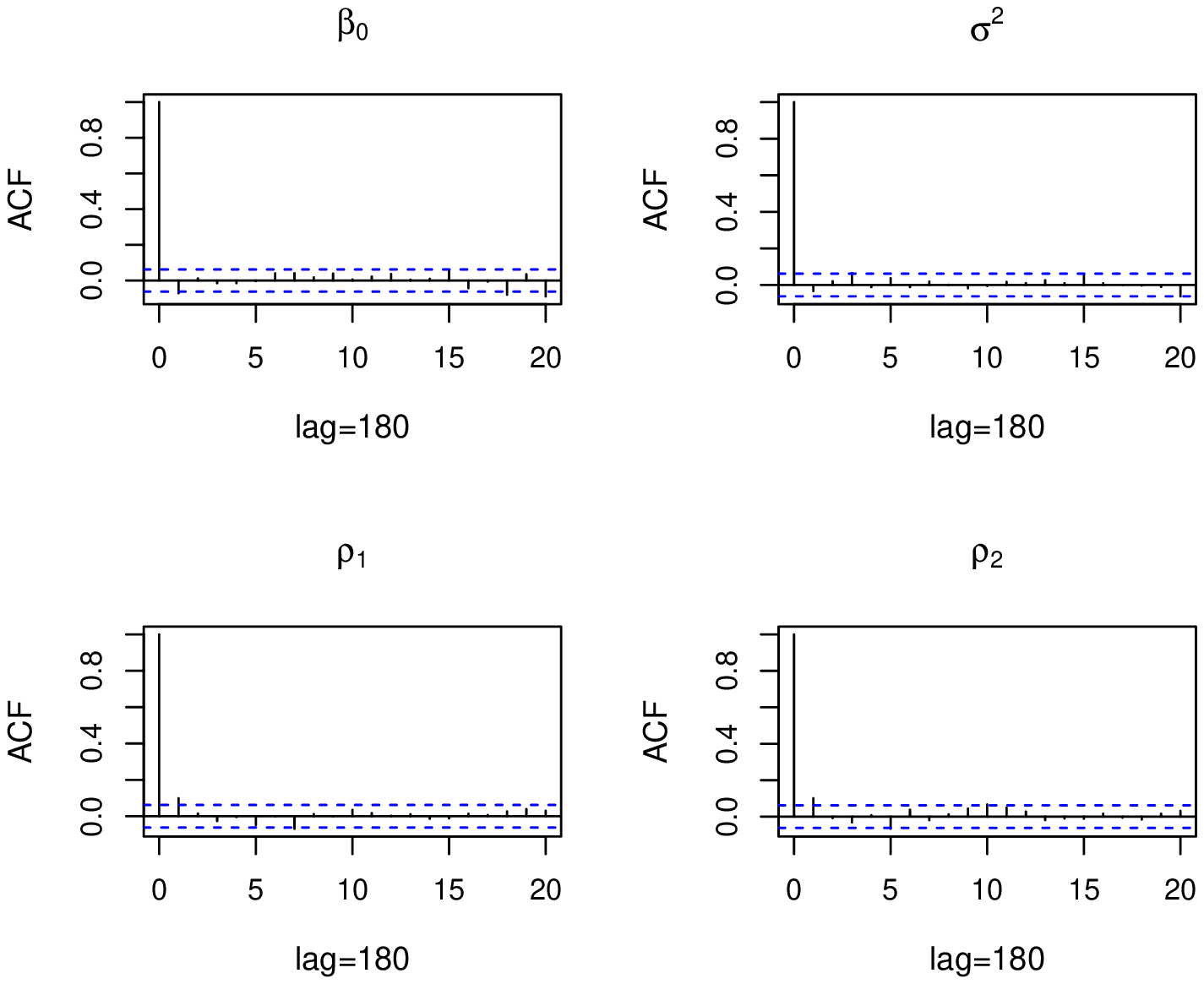} 
		\caption{\textit{Top}: Geweke plots and ACF functions of the subsamples used to compute the parameters estimates for CLR-AR(1). \textit{Botton}: Geweke plots and ACF functions of the subsamples used to compute the parameters estimates for CLR-AR(2).} 
		\label{fig:gwk_acf}  
              \end{figure}

              \begin{figure}[h!] 
            \includegraphics[width=6cm]{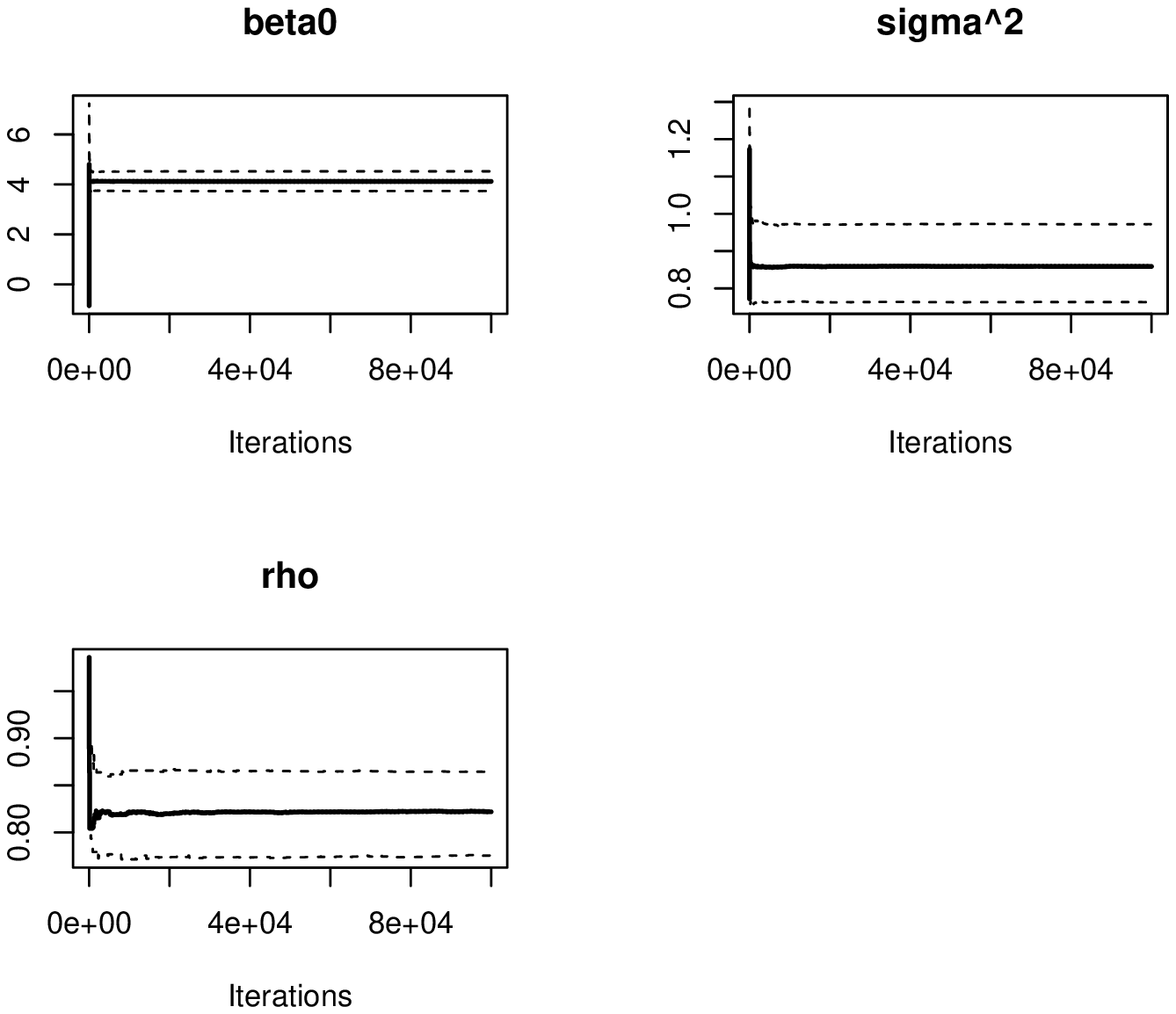} 
		\includegraphics[width=6cm]{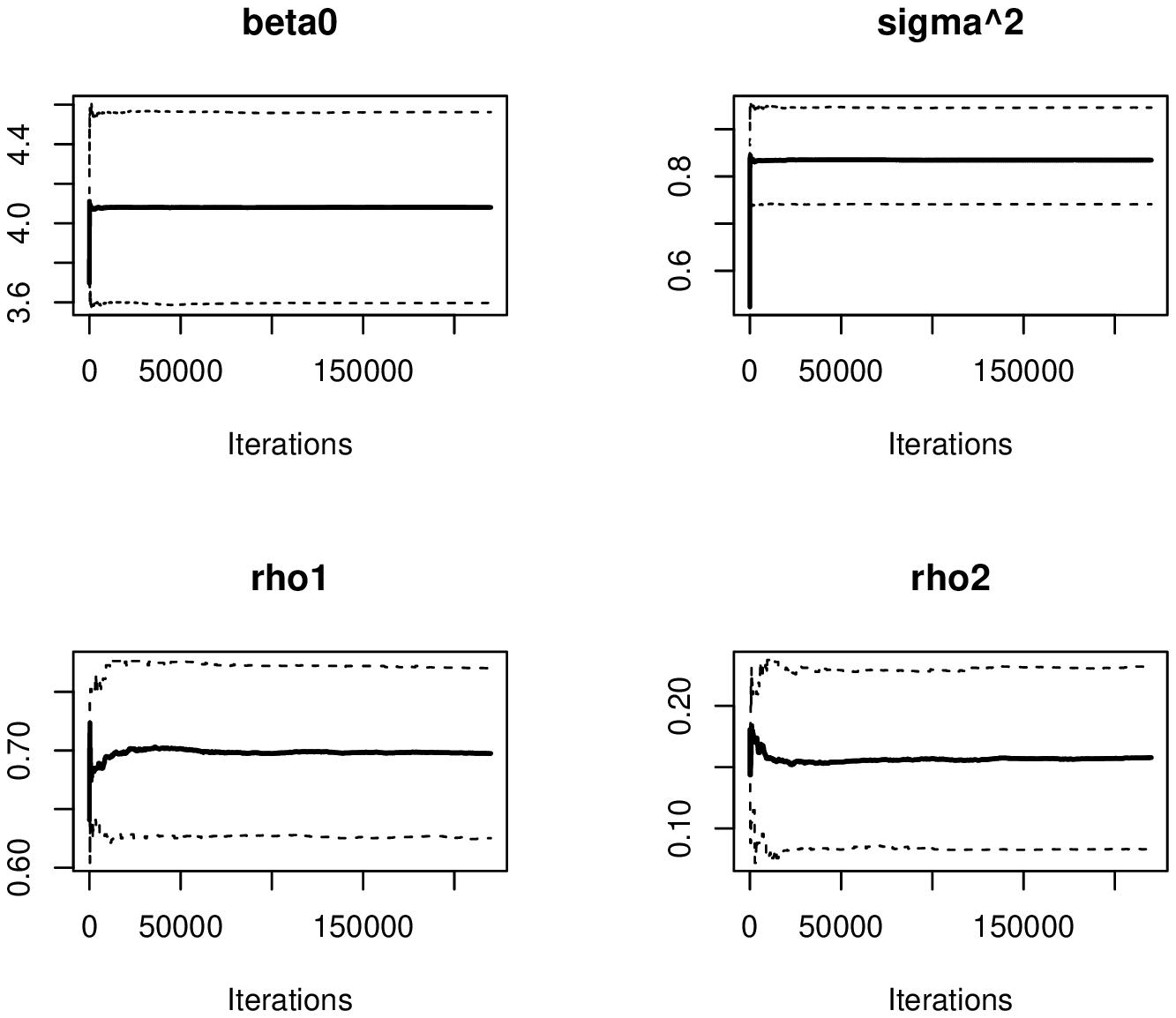} 
		\caption{Evolution of 1st and 3rd quantiles corresponding MCMC outputs,\textit{top}:  AR($1$) model; \textit{bottom}: AR($2$) model.} 
		\label{fig:ch5_pd_cmp_app} 
\end{figure} 
              
\end{document}